\documentclass{aa}  

\usepackage{graphicx}
\usepackage{xcolor}
\definecolor{RUGred}{RGB}{150,0,0}
\usepackage{txfonts}
\usepackage[colorlinks=true]{hyperref}
\hypersetup{
    colorlinks=true,
    linkcolor=RUGred,
    filecolor=RUGred,      
    urlcolor=blue,
    citecolor=blue
    }
    
\usepackage{float}
\usepackage{subcaption}
\usepackage{textpos}
\usepackage{wrapfig}
\usepackage[export]{adjustbox}

\usepackage{makecell}

\newcommand{\teff}{\rm T_{eff}}
\newcommand{\logg}{\log(g)}
\newcommand{\feh}{\rm [Fe/H]}

\newcommand{\apoteff}{\textsc{TEFF}}
\newcommand{\apologg}{\textsc{LOGG}}
\newcommand{\apofeh}{\textsc{FE\_H}}
\newcommand{\apomh}{\textsc{M\_H}}

\begin{document}

   \title{The metal-poor tail of the APOGEE survey}

   \subtitle{II. Spectral analysis of Mg and Si in very metal-poor APOGEE spectra}

   \author{M. Montelius
          \inst{1}\fnmsep\thanks{\email{montelius@astro.rug.nl, a.angrilli-muglia@herts.ac.uk}},
          A. Angrilli Muglia
          \inst{1}\fnmsep\thanks{The first and second author contributed equally to the work.} 
          \and
          E. Starkenburg\inst{1}
          \and
          C. Kobayashi\inst{2}
          \and
          A. Ardern-Arentsen\inst{3}
          \and 
          F. Gran\inst{4}
          \and
          P. Jablonka\inst{5}
          \and
          N. Martin\inst{6,7}
          \and
          J. Navarro\inst{8}   
          \and
          F. Sestito\inst{2}
          \and
          K. A. Venn\inst{9}
          \and
          S. Vitali\inst{10}}

   \institute{
        Kapteyn Astronomical Institute, University of Groningen, Landleven 12, NL-9747 AD Groningen, the Netherlands
        \and 
        Centre for Astrophysics Research, Department of Physics, Astronomy and Mathematics, University of Hertfordshire, College Lane, Hatfield AL10 9AB, UK
        \and 
         Institute of Astronomy, University of Cambridge, Madingley Rd, Cambridge CB3 0HA, UK
        \and 
        Université Côte d’Azur, Observatoire de la Côte d’Azur, CNRS, Laboratoire Lagrange, Bd de l’Observatoire, CS 34229, 06304, Nice Cedex 4, France 
        \and 
        Laboratoire d’astrophysique, École Polytechnique Fédérale de Lausanne (EPFL), 1290 Sauverny, Switzerland
        \and 
        Université de Strasbourg, CNRS, Observatoire astronomique de Strasbourg, UMR 7550, F-67000 Strasbourg, France
        \and 
        Max-Planck-Institut für Astronomie, Königstuhl 17, D-69117 Heidelberg, Germany 
        \and 
        CIfAR Senior Fellow and Professor. Department of Physics and Astronomy, University of Victoria, Victoria, BC, Canada V8P 5C2
        \and
        Department of Physics and Astronomy, University of Victoria, PO Box 3055, STN CSC, Victoria, BC V8W 3P6, Canada
        \and 
        Instituto de Estudios Astrof\'isicos, Facultad de Ingenier\'ia y Ciencias, Universidad Diego Portales, Av. Ej\'ercito Libertador 441, Santiago, Chile}

   \date{Received 11 December 2025 / Accepted 13 January 2026 }

  \abstract
   {Very metal-poor stars are important tracers of the early chemical evolution history of the Milky Way. Infrared H-band spectroscopic surveys, such as APOGEE, are less affected by extinction in the more dust-obscured environments of our Galaxy. However, H-band spectra contain very limited spectral information for stars at the most metal-poor tail ([Fe/H] < -2.5) because the available Fe lines in FGK stars in this wavelength range are weak.}
   {The first paper in this series successfully identified a sample of 327 very metal-poor stars (with [Fe/H] $< -2$) from the APOGEE database, 289 of which are on the red giant branch. The spectra of these stars were not properly analysed by the APOGEE main pipeline because they are very metal poor. In this work, we measure metallicities for these stars using the abundances of the elements Mg and Si.}
  {We demonstrate that the absorption lines of the elements Mg and Si are of good quality despite the challenging combination of (low) metallicity, wavelength regime, spectral resolution, and signal-to-noise ratios available for these spectra. A specialised pipeline was designed to measure the abundance of Mg and Si in APOGEE spectra and  yielded a robust estimate of the overall metallicity. In order to provide reliable measurements, we tested three different sets of assumptions for Mg and Si enhancement. }
 {We present Mg and Si abundances as well as overall metallicities for 327 stars, all of which had previously gotten null values from the main APOGEE pipeline for either the calibrated [M/H] or [Fe/H]. The typical uncertainties for our measurements are 0.2 dex. We found five stars in our sample with unusual [Si/Mg] abundances higher than 0.5, and we connect this signature to globular cluster stars, and this might be related to specific supernova events. Our data suggest a concentration of high [Si/Mg] stars in the Sextans dwarf galaxy. Other dwarf galaxies are found to agree well with results in the literature. }
   {Our derived metallicities range between $-3.1\leq[M/H]\leq-2.25$, thereby pushing the metal-poor tail of APOGEE results down by 0.6 dex.  }


   \maketitle

\section{Introduction}
The outer layers of the atmosphere of a low-mass star remain a snapshot of the environment in which it formed. This opens the possibility to use metal-poor stars as a fossil record through which we can witness a long-gone landscape of chemical evolution \citep[see for a review][and references therein]{beers_2005,frebel_near-field_2015, bonifacio_2025}. By using spectral analysis to measure the abundance ratios of different chemical elements, the chemical evolution history in a given environment at a given point in time can be constrained. 

The numerous spectral lines of Fe I and Fe II that are present in the optical wavelength region have made it common practice to use [Fe/H]\footnote{This notation is defined as [A/B] = $\log (N_\mathrm{A}/N_\mathrm{B})_* - \log (N_\mathrm{A}/N_\mathrm{B})_\odot$, with $N_\mathrm{A}$ and $N_\mathrm{B}$ being the number densities of elements A and B.} as a proxy for the overall metallicity. However, for the most metal-poor stars, which trace the earliest chemical evolution, the Fe lines can become very weak \citep[see e.g. Fig. 1 of][]{frebel_2010}. This makes them difficult to detect and measure, especially in spectra with low or intermediate spectral resolution and low signal-to-noise ratios (S/N\footnote{We use the S/N reported by APOGEE, which is calculated per pixel across the full wavelength range of APOGEE, including all pixels not affected by telluric absorption or sky lines \citep{Nidever2015}.}). To measure metallicities in the lowest metallicity regimes, the strongest absorption lines in the spectra are therefore often targeted instead. In the optical wavelength regime, the strongest metal absorption lines in FGK-type stars include the Mg I triplet and the Ca II triplet, but even more prominently, the very strong Ca II H\&K lines \citep[e.g.][, and their discussion on the lowest observable metallicity]{frebel_2013}.
However, to place the metallicities derived from these strong lines on the same scale as directly measured Fe lines, (implicit) assumptions on the nucleosynthesis in stars and chemical evolution are needed \citep[e.g.][]{burbidge_1957, Nomoto2013ARA&A..51..457N}. 

Mg and Ca belong to the group of $\alpha$ elements, which are thought to mainly originate from core-collapse supernovae (CC SNe), while iron-peak elements, such as Fe, are thought to form largely in supernovae type Ia (SNe Ia). 
While CC SNe continuously form both $\alpha$ and iron-peak elements at a steady pace, iron-peak elements form much more rapidly after the onset of SNe Ia \citep[see more detailed discussion in e.g.][]{nissen_high-precision_2018}.
In a plot of [$\alpha$/Fe] versus [Fe/H], this is visible as a plateau in the [$\alpha$/Fe] values for the most metal-poor stars, representing the ratio before the onset of SNe Ia, broken by a knee at which the ratio starts to decrease due to the rapid Fe production from SNe Ia \citep[this type of explanation, relating stellar lifetimes and chemical abundance ratios goes back to pioneering work by][]{tinsley_1979}.

The exact ratio at which this plateau is found is different for each $\alpha$ element and for each stellar system, with typical values ranging from 0.3-0.6 \citep[e.g.][]{kobayashi_galactic_2006, matteucci_modelling_2021, zhang_determining_2024, velichko_alpha-element_2024}. Additionally, some $\alpha$ elements receive contributions from different pathways. For instance, Si and Ca are (also) produced in SNe Ia, while Mg is not \citep{kobayashi_origin_2020}. Pair-instability supernovae (PISN) would produce a large amount of Ca and Si compared to Mg \citep{takahashi_2018_stellar}. The explosion energy of SNe type II also affects the ratio of Si and Mg, with higher amounts of Si released in the more energetic hypernova \citep[e.g.][]{Heger2010}. Measuring the ratio of [Si/Mg], we can therefore trace the chemical enrichment history of a stellar system.

Several current surveys, as well as future instrumentation, observe in (near-) infrared wavelengths instead of in the optical wavelength regime \citep[e.g. NIRSpec, MOONS, ANDES\footnote{Near-Infrared Spectrograph, Multi-object Optical and Near-IR spectrograph, and ArmazoNes high Dispersion Echelle Spectrograph} -][]{ferruit_nirspecJWST_2012,cirasuolo_moons_2020, marconi_andes_2024}. In addition to obvious advantages for high-redshift studies, this wavelength regime also benefits studies of different environments within the Milky Way because infrared light can more easily penetrate the gas and dust that obscures the Galactic bulge and disc, enabling higher S/N with shorter integration time. To date, the Apache Point Observatory Galactic Evolution Experiment \citep[APOGEE,][]{majewski_apache_2017, abdurrouf_seventeenth_2022}, is the largest stellar spectroscopic survey and observed in the infrared H band. However, for H-band spectra, the metallicity at which Fe lines become undetectable at a typical S/N is significantly higher than the limit for optical spectra. This is due to the atomic structure, with the higher-wavelength lines typically originating from more sparsely populated high excitation energy states \citep[e.g.][]{smith_apogee_2021}. While for large optical spectroscopic surveys, the most metal-poor tail is often carefully analysed, usually by an additional targeted pipeline beyond the main analysis pipeline to identify and analyse very metal-poor stars specifically \citep[e.g.][]{SEGUE_trashcan2016, Li2018, Matsuno2022, LAMOST_trashcan2023, Hou2024, Viswanathan2024}, no such re-analysis has been done so far for APOGEE because no Fe lines are available in stars with [M/H]\footnote{[M/H], or metallicity, is sometimes used interchangeably with [Fe/H]. In this paper we use [M/H] to represent the same quantity denoted as $\apomh$ by APOGEE, derived by full spectrum fitting.} < $-2.5$.

In the first paper of this series (Montelius et al., 2025, hereafter Paper I), we have identified a sample of very metal-poor stars from APOGEE by using its calibrated $\apoteff$ and $\apologg$ and the lack of either calibrated $\apomh$ or $\apofeh$. This selection furthermore included a minimum S/N of 30 for the spectrum to ensure that the lack of measured metallicity is not due to a low S/N and a minimum probability of being a star of 95\%  in the discrete source classifier from \citet{delchambre_gaia_2023} to remove non-stellar sources. Additionally, the spectra were inspected visually, and any spectra with emission lines (likely young stellar objects) or spectra with key absorption lines too close to the spectral gap in the APOGEE wavelength coverage (as discussed in more detail in Sect. \ref{sec:methods}) were removed from the sample. A literature comparison, shown in Paper I Sect. 3, showed that the resulting sample has almost no higher-metallicity interlopers. In particular, all stars on the red giant branch (RGB) overlapping with the literature were found to have [Fe/H] $< -2$ without outliers. For candidates in other parts of the Kiel diagram (main sequence, sub-giant branch, and horizontal branch) the selection appeared less clean, but the literature overlap was also much sparser in these regimes.

In this paper, we analyse the APOGEE spectra for the sample defined in Paper I. In Section \ref{sec:data}, we describe the sample and the lines we used for the analysis. Section \ref{sec:methods} describes the method we developed to measure abundances for these stars, and the assumptions we used to convert them into overall metallicities. Section \ref{sec:results} describes the results of this analysis and compares the [Si/Mg] abundances in stars from different substructures in our sample. We also include a comparison with literature studies for this abundance ratio. The last section summarises and discusses our main results and explores potential applications of this method in the future. \par

\section{The sample of very metal-poor stars from APOGEE}
\label{sec:data}
Continuing from the results in Paper I, we took two subsamples as input for our analysis. First, we took all very metal-poor candidates that fall along the red giant branch (RGB) in the Kiel diagram of APOGEE stellar parameters (see for an illustration of the cuts adopted, Paper I, Fig. 6). This subsample of 289 stars, hereafter the high-confidence sample, was studied kinematically in Paper I. Additionally, we also analyse 38 stars that form the lower-confidence sample due to their placement on the Kiel diagram where they trace either the main-sequence turn-off, the horizontal branch, or the sub-giant branch. 
We use APSTAR spectra downloaded from \url{https://data.sdss.org/sas/dr17/apogee/spectro/redux/dr17/stars/} with wget, using the \textsc{APSTAR\_ID} and \textsc{FILE} columns to find the correct files. If multiple spectra were found, only the highest S/N was kept.

\section{Spectral analysis method}\label{sec:methods}

\subsection{Lines and wavelength segments for the analysis}

Very low-metallicity stars have few metallic absorption lines in the infrared. The only lines suitable for analysis that could be identified from visual inspection of the spectra correspond to the $\lambda$ $15740.7$ $\AA$, $ 15748.9$ $\AA$, and $ 15765.8$ $\AA$ Mg I lines and the $\lambda$ $15888.4$ $\AA$, and the $15960.0$ $\AA$ Si I lines. We note that the APOGEE spectra were carefully inspected for other features, including known carbon and aluminium atomic lines \citep[see e.g.][]{HbandC-1993A&AS...99..179H, HbandAl-1990JPCRD..19..119C}, as well as potential molecular lines, but no other analysable feature was found at these metallicities.

 The three magnesium lines mentioned above are within 30 $\AA$ of each other, while the silicon lines are only 70 $\AA$ apart. Given this proximity, we choose to perform our analysis on two segments, one for Mg and one for Si. This allows us to normalise the spectrum locally. We define the region between $ 15720$ $\AA$ and $ 15790$ $\AA$ as the magnesium segment, while we denote the region between $ 15860$ $\AA$ and $ 16000$ $\AA$ as the silicon segment. We note that the silicon segment is close to a gap in the APOGEE spectral coverage that runs from $\sim15810$ $\AA$ to $\sim15858$ $\AA$. Additionally, while the $15888.4$ $\AA$ silicon line is one of the intrinsically strongest lines in the infrared, it is in the wings of a hydrogen line. Nevertheless, given the relative shortness of the segment and the absence of other absorption lines, we define wavelength ranges around each line to be used as continuum masks. Using these masks, we fit a first-degree polynomial to normalise each segment. Our pipeline automatically excludes regions affected by incorrect telluric or cosmic ray removal, bad pixels, or any other artifact the spectra might present.

\subsection{Atomic and solar data}\label{sec:ll}
In line formation, the line intensity is proportional to both the number density of a certain element, and to the log(gf) value (where g is the statistical weight and f is the oscillator strength). When measuring abundances, a shift in abundance is equivalent in magnitude to an offset in log(gf) \citep[see for instance][]{rutten_radiative_2003, gray_observation_2008}, underlining the importance of having reliable atomic data when measuring individual absorption lines. In the near-infrared wavelength regime that the APOGEE survey probes, the line list and their log(gf) values are less well tested than in the optical wavelength regime where stellar spectroscopy has a longer and richer history (although we note there are some interesting attempts to also explore weak and blended species in the APOGEE wavelength regime in \citealt{Hayes2022}). 

As input for our pipeline, we use a line list from \cite{2022A&A...665A.135M} based on the VALD3 database \citep{VALD2019ARep...63.1010P}. We have updated the linelist with updated calculations of log(gf) from \cite{pehlivan_rhodin_experimental_2017} for Mg lines and \cite{pehlivan_rhodin_accurate_2024} for Si lines. Their values are reported in Table \ref{tab:lines}. 
Both sources are cited in APOGEE's linelist \citep{smith_apogee_2021}, with the Si values attributed to the version in preparation of \cite{2018PhDT.......134P}.
The reference solar values we use throughout this work are taken from \cite{asplund_chemical_2009}. 

\subsection{Stellar parameters}
\label{sec:params}
Even though ASPCAP does not converge to calibrated metallicities for the stars in our sample, meaning there are no (calibrated) metallicity values available in the catalogue, the calibrated values for effective temperature ($\apoteff$) and surface gravity ($\apologg$) are available. 

The largest overlapping sample with stellar parameters (140 stars), is table 2 from \cite{Andrae2023b}. These values are derived from Gaia XP spectra using machine learning trained partially on APOGEE data. Despite not getting a calibrated metallicity, 24 of the stars in our sample are used in the training sample. By comparing the two samples, we therefore cannot get an independent evaluation of our stellar parameters, but we can get a confirmation of whether the values for our stars are in line with what is expected for APOGEE stars.
For $\apoteff$ there is a mean offset of 21 K and a standard deviation of 109 K, for $\apologg$ the offset is -0.01 dex and the standard deviation is 0.24 dex. This is in line with what we expect for the values from \cite{Andrae2023b}. 

28 stars have stellar parameters from the SAGA data base of metal-poor stars \citep{suda_stellar_2008}\footnote{Using data from \cite{2002ApJ...567.1166A, 2004ApJ...607..474H, 2004PASJ...56.1041S, 2005ApJ...632..611A, 2009A&A...494.1083A, 2009ApJ...705..328K,  Kirby2010ApJS..191..352K, 2010A&A...524A..58T, zhang_hamburgeso_2011, 2012A&A...537A.118R, 2015ApJ...807..171J, 2015ApJ...798..110L, 2017ApJS..230...28N, 2018ApJ...868..110S, 2018A&A...611A..30S, 2019A&A...622L...4A, li_four-hundred_2022}}. We use the recommended data set from April 2021, combining [Fe/H] and [M/H] measurements as in Paper I. For $\apoteff$ the mean offset with the calibrated APOGEE parameters is 161 K with a standard deviation of 165 K, with an offset and standard deviation of 0.38 and 0.45 for $\apologg$.  There appears to be a metallicity dependence in the offsets, with the metal-poor stars slightly offset to higher temperatures in APOGEE compared to SAGA. As these offsets are within one standard deviation and the sample of offset stars is small ($\sim5$) we do not expect this to have a major impact on our analysis. For the lower-confidence sample, only two stars from this sample have stellar parameters from the SAGA database. Both are turnoff stars, and are therefore among the hottest stars in the sample, and not representative of all of the lower-confidence stars. While the first star matches the SAGA $\teff$ down to 20 K and has an offset in $\logg$ of -0.8, the second has a 300 K offset in $\teff$ and only -0.12 in $\logg$. In general we expect the lower-confidence sample to have larger uncertainties in their stellar parameters, especially as many of them are hotter, and therefore are less likely to have any metallic lines in their spectra. 

The uncalibrated stellar parameters from APOGEE are somewhat better aligned with the SAGA parameters, with an offset of -96 K for $\teff$ and -0.27 for $\logg$ and similar standard deviations. As our pipeline described in Sect. \ref{sec:flowchart} often failed to converge with the uncalibrated parameters we opted to use the calibrated parameters.

\begin{table}
\caption{Wavelengths and log(gf) values for the lines we analysed in section \ref{sec:methods}, along with their fine-structure components.}
\label{tab:lines}  
\centering   
\begin{tabular}{c c c}     
\hline\hline 
$\lambda$ ( $\AA$) & Element & log(gf)\\
\hline
15740.716 & Mg I & -0.223\\ \hline
15748.886  & Mg I & -0.348\\ \hline
15748.988 & Mg I & 0.129\\ \hline
15765.645 & Mg I & -1.524\\ \hline
15765.747 & Mg I & -0.348\\ \hline
15765.842 & Mg I & 0.400\\ \hline
15888.409 & Si I & -0.001\\ \hline
15960.060 & Si I & 0.170\\ \hline
\end{tabular}
\end{table}

\subsection{Pipeline for the abundances}\label{sec:flowchart}
Elemental abundances are measured using the Python version of the Spectroscopy Made Easy (SME) code (\citealt{valenti_spectroscopy_1996}, \citealt{valenti_spectroscopic_2005}, \citealt{piskunov_spectroscopy_2017}, \citealt{wehrhahn_pysme_2023}). A 1D LTE MARCS \citep{gustafsson_grid_2008} grid of atmosphere models was adopted, along with non-local thermodynamic equilibrium (NLTE) departure coefficients for both Mg and Si \citep[following][]{amarsi_galah_2020}. We adopt spherical geometry for the model atmospheres where possible (in particular for the majority of our targets in the VMP\_RGB sample with $0.5 < $log$(g) < 3.5$), only employing 1D geometry to fill in any gaps in parameter space.  

\begin{figure}[h]
    \centering
    \includegraphics[width=\linewidth]{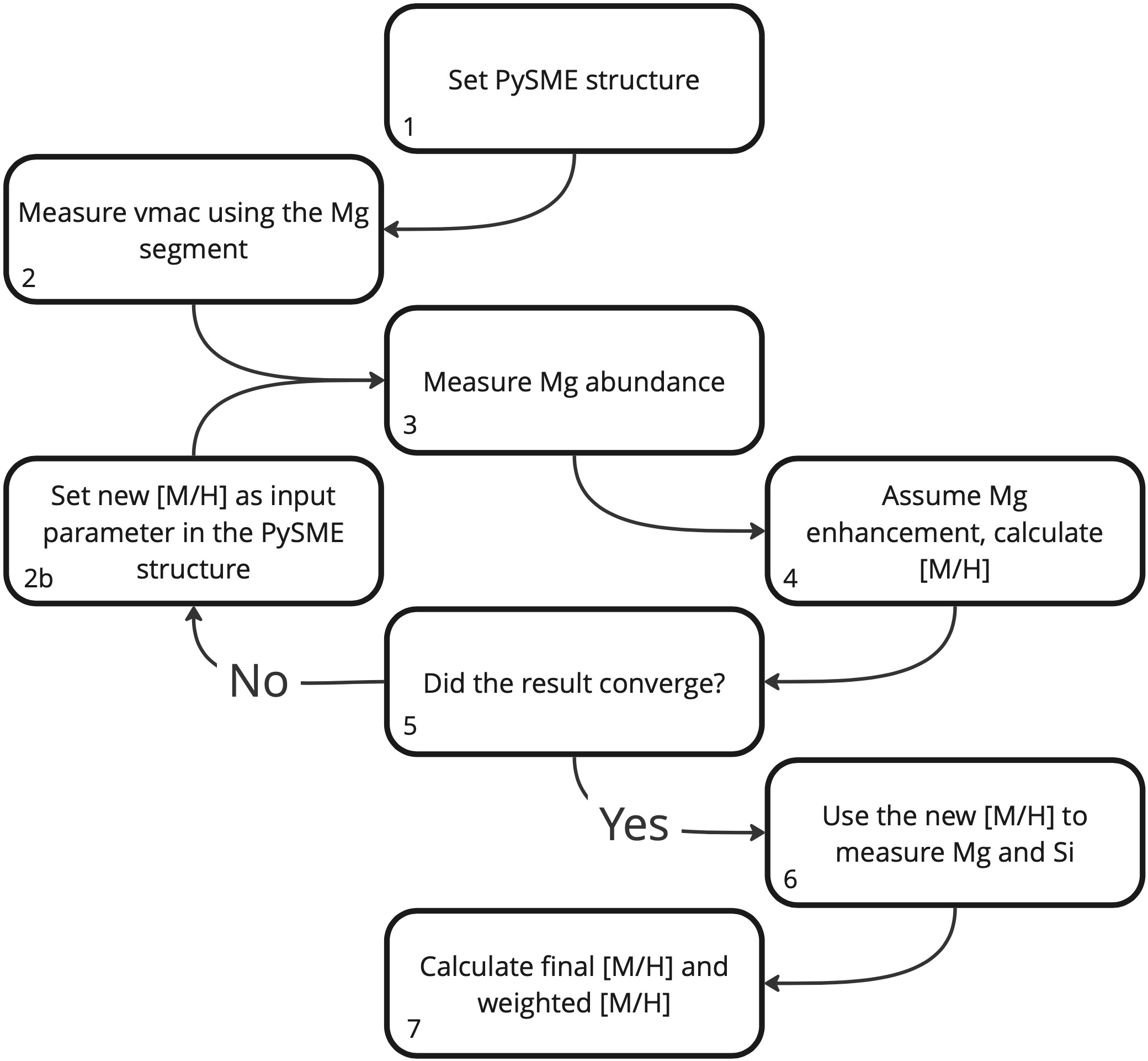}
    \caption{Flowchart of the pipeline, see Sect. \ref{sec:flowchart} for details.}
    \label{fig:flowchart}
\end{figure}

The pipeline used in this work to measure abundances and derive a value for metallicity was written in Python and follows the multiple steps that are schematically shown in Fig. \ref{fig:flowchart}. In the following we will follow the step structure shown in this figure to explain the procedure in more detail.

\begin{enumerate}
\item{For the analysis of each star, the APOGEE main pipeline effective temperature, surface gravity, and microscopic velocity were used as inputs to PySME. Our initial assumption for metallicity is set to APOGEE's uncalibrated metallicity ($[M/H] \sim $-2.5 for the vast majority of our sample).}
\item{The macroturbulence velocity is measured from the Mg segment. For this purpose, the Mg segment is preferred over the Si segment, because it typically has a higher S/N, is further in wavelength from the APOGEE spectral gap, and the lines are closer to each other in wavelength making the continuum derivation more precise. However, macroscopic velocities did not always converge for all of our objects. If no convergence is reached and the value returned is the grid edge, we manually set the value to $\overline{V}_{mac} = 10.5$ km/s. This value was calculated by averaging the macroturbulence velocity results for a calibrated and cleaned\footnote{For this reference dataset we removed all the stars with less than 70 S/N and less than three visits, and filtered by the following \textsc{STARFLAGS} and \textsc{ASPCAPFLAGS}: \textsc{BAD\_PIXELS}, \textsc{VERY\_BRIGHT\_NEIGHBOR}, \textsc{LOW\_SNR}, \textsc{STAR\_BAD}, \textsc{M\_H\_BAD}, \textsc{TEFF\_WARN}, and \textsc{LOGG\_WARN}. We filtered these stars further by performing 5-$\sigma$ clippings on the macroscopic velocity.} set of APOGEE stars with metallicities between -2 and -2.5.}
\item{We then calculate the Mg abundance for each of the program stars using the Mg segment.}
\item{To convert the measurement of [Mg/H] to a general [M/H] consistent with the scale used by the APOGEE main pipeline, an assumption has to be adopted on [Mg/M] for stars in our sample. Several assumptions for this value have been tested and compared to literature data from high-resolution studies in the optical wavelength regimes (see Sect. \ref{sec:Mchoices}). As a result of these tests we have adopted [Mg/M] = [Si/M] = 0.4 in this work. We note that this there are some stars in our sample that are clearly anomalous in [Si/Mg] and for which this general assumption will not be valid. These are discussed separately in Sect. \ref{sec:SiMgCauses}. }
\item{The resulting [M/H] value is compared to the [M/H] value currently adopted in Step 2, and is regarded as converged if the difference between the two values is less than 0.1 dex after which the code will proceed to Step 6. If no convergence is reached, the new [M/H] is used as an input for Step 2 and Steps 2 until 5 are repeated. In the vast majority of the cases, only one iteration (after the initial guess) is needed to reach this level of convergence. }
\item{With the updated [M/H] value as input for the PySME structure, the Mg abundance is calculated again from the Mg segment and the Si abundance is calculated from the Si segment.}
\item{Using the assumption of [Mg/M] = [Si/M] = 0.4 as mentioned above, both measurements of [Si/H] and [Mg/H] are combined to derive a final value for [M/H]. In this final value, the [M/H] is the weighted average of the [M/H] value derived from the Mg and Si segments separately. The uncertainties on the measurements (discussed in detail below) are used as weights in this calculation.}
\end{enumerate}

For the vast majority of our stars, the pipeline as described above could be fully followed. However, in some cases the [Mg/H] or [Si/H] measurement did not converge, due to the segment being too noisy, containing damaged pixel measurements in critical places, or the abundance of one of the elements being too low. When this occurred, the first five steps of the pipeline were still followed -- if needed by replacing Mg with Si in Steps 2, 3 and 4 (when the Mg segment was not converging). In these cases, the final metallicity for the star was thus derived using the one available element only. 

\begin{figure*}
    \centering
    \includegraphics[width=1\linewidth]{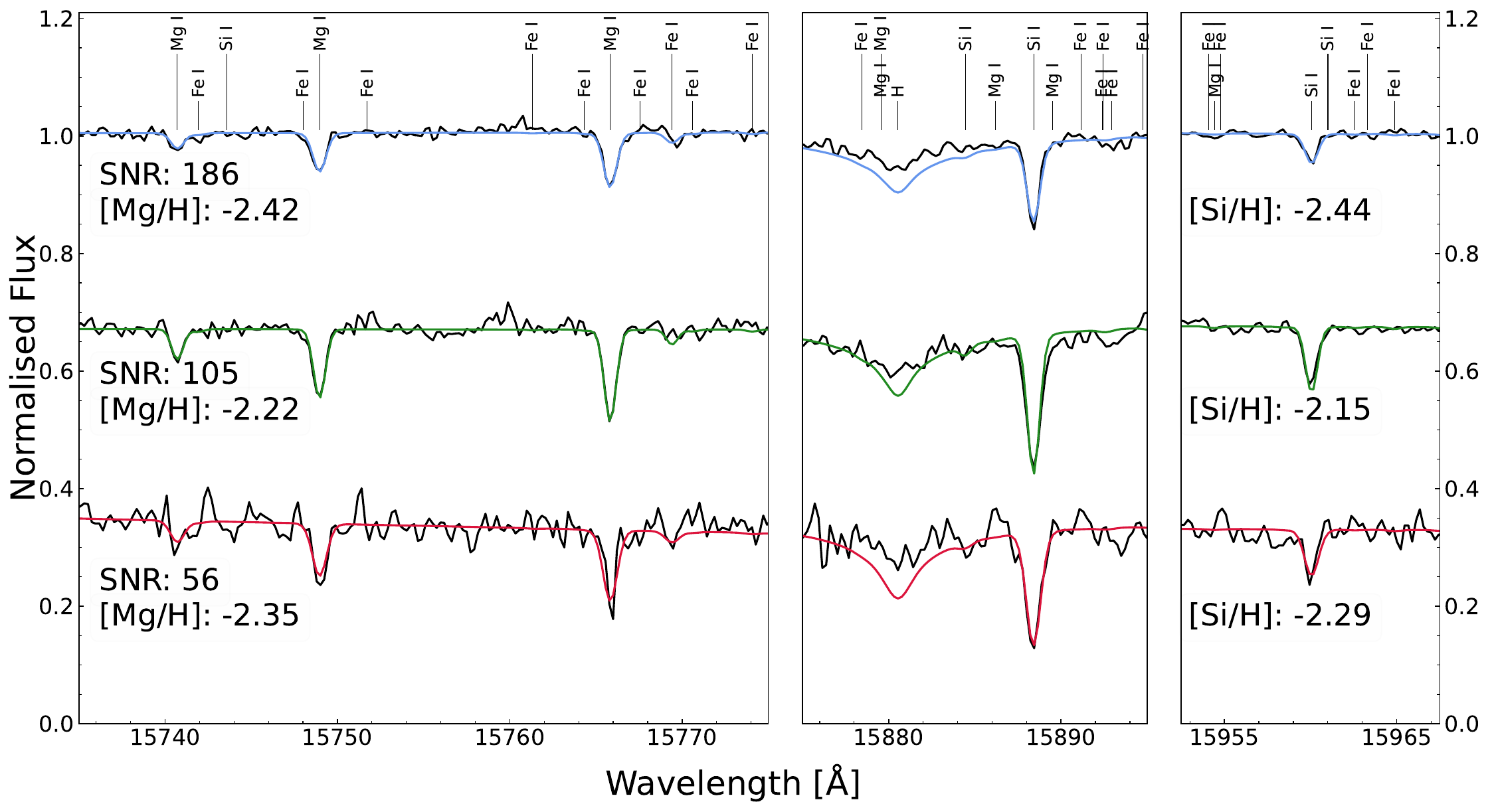}
    \caption{Example synthetic fits to some of the program spectra, the spectra are artificially offset to make all three visible. The spectra are chosen to represent the median S/N, and the upper and lower quantiles.} 
    \label{fig:spec_SNR_fit}
\end{figure*}

Figure \ref{fig:spec_SNR_fit} shows three spectra chosen to represent the median S/N for the high-confidence sample (S/N = 104), and the 16th and 84th percentiles (S/N = 58, 186). The synthetic spectra fit to each spectra to measure abundances is also shown together with the derived abundance. The identification of atomic lines shown at the top of the plot are taken from the Arcturus atlas of \cite{ArcturusAtlas}. The main discrepancy between observed and synthetic spectra is seen at the hydrogen line at 15880 \AA. This is due to the use of APOGEE's calibrated temperatures, and not to the uncalibrated values taken directly from the spectra. For the high-confidence sample, the calibrated values are 250 K hotter than the uncalibrated ones on average, which would explain the strengthened hydrogen lines. At lower S/N the weaker lines (Mg I at 15740 \AA\ and Si I at 15960 \AA) can be difficult to analyse, especially for the lowest metallicity stars. 

\subsection{Choice of the [Mg/M] and [Si/M] values}\label{sec:Mchoices}
As noted in Sect. \ref{sec:flowchart}, our metallicity estimate requires an assumption for [Mg/M] and [Si/M] in our sample. We stress that this is not at all uncommon at these very low metallicities, as the Fe lines get weaker before some of the stronger lines of $\alpha$-elements, metallicities in this regime are often based on these strongest absorption lines (in the optical often the Ca II H\&K lines, the Mgb triplet, or the Ca II triplet for instance). While in this work, we adopt [Mg/M] = [Si/M] = 0.4, we have additionally tested the effect of two different assumptions for the enhancements:
\begin{itemize}
    \item Using the same quality cuts used to find macroscopic velocity, that is, filtering using flags and 5-$\sigma$ clippings (see Sect. \ref{sec:methods}), we calculated the average [Mg/M] and [Si/M] of stars with metallicities between -2 and -2.5 from the APOGEE sample, resulting in [Mg/M] = 0.36  and [Si/M] = 0.28. As APOGEE presents abundances with the most commonly used nomenclature [X/Fe], [X/M] has been calculated using the general formula [X/M] = [X/Fe] + [Fe/H] - [M/H].
    \item Theoretical models from \cite{kobayashi_origin_2020} predicts that at [M/H] = -2.5 the approximate enhancements for the analysed elements are [Mg/M] = 0.45 and [Si/M] = 0.545. 
  \end{itemize}

The three enhancement assumptions were compared to previously measured metallicities from SAGA, the reanalysis of LAMOST DR10 spectra used in Paper I \citep[following][]{AnkeLAMOST2025}, and the Pristine survey with photometrically derived metallicities \citep{martin_pristine_2023} by calculating their metallicity differences, that is, $\Delta [M/H] = [M/H]_{\rm this\,work,\,weighted} - [M/H]_{\rm survey}$. The mean deviations for each assumption are: $\overline{\Delta [M/H]_{APO}} = 0.069$, and $\overline{\Delta [M/H]_{Kob}} = -0.107$ and $\overline{\Delta [M/H]_{0.4}} = -0.012$. As our simplest assumption, [Mg/M] = [Si/M] = 0.4, yielded the best results in comparison with these independent datasets, we have kept this assumption for the purpose of this work. 

For stars with either high or low values of [Si/Mg] this assumption is unlikely to hold. Estimating the effect this has on the metallicity would require further assumptions on whether changes arise from enhancement or depletion of either or both of the elements. To keep the method as general as possible, we instead add an extra term to the uncertainty on the metallicity of $\rm |[Si/Mg]|/3$.

\subsection{Estimation of the uncertainties}
For the final uncertainty of each abundance measurement, we take into account three main factors: uncertainties on the stellar parameters, uncertainties in the continuum fit, and the uncertainties on the abundance measurement. 

While APOGEE's pipeline ASPCAP returns uncertainties for \apoteff\ and \apologg, we note that these are standard deviations based on multiple observations, and can therefore be unrealistically low (down to 10K for \apoteff\ and 0.01 for \apologg\ for our sample). 
Instead of using these values, we adopt an uncertainty for $\teff$ of 200 K based on the comparison of stellar parameters with the SAGA database described in Sect. \ref{sec:params}.
Because uncertainties on $\teff$ and $\logg$ are highly correlated we re-calculate the $\logg$ of the star based on this uncertainty in $\teff$ and investigate its effect on the derived abundances. In practice, every star is mapped to a corresponding phase of a MIST \citep{paxton_modules_2011, dotter_mesa_2016, choi_mesa_2016} stellar isochrone selected for metallicities between -2 and -3 in 0.5 increments, and with an age of 10 Gyr. Each isochrone is pre-divided into distinct evolutionary phases. This is of importance, as for a main sequence star a higher temperature corresponds to a marginal decrease in gravity, while, in contrast, for RGB stars a higher temperature corresponds to a relatively steep increase in gravity. For an isochrone with metallicity \(i\) and phase \(j\), each star was matched to the nearest point on the closest isochrone based on its initial stellar parameters, denoted as \(s_{ij}\). A pair of points, \(l\) and \(m\), were selected such that the temperature difference \(\Delta T_{ij} = T_{ijl} - T_{ijm}\) = 200 K. A corresponding gravity difference \(\Delta \log(g)_{ijn} = \log(g)_{ijl} - \log(g)_{ijm}\) was determined by placing the star again on the same isochrone. These adjusted parameters of $\teff$ and $\logg$ were then input into the PySME pipeline discussed in Sect. \ref{sec:flowchart}, resulting in new estimates for [X/H]. The impact of the uncertainty over atmospheric parameters was finally calculated using a change in $\teff$ of $+200$ K (upper) and $-200$ K (lower)\footnote{Occasionally, adding or subtracting $200$ K and adjusting gravity accordingly would take stars outside the utilised atmospheric grids, meaning it was not possible to calculate the error on either side. In such cases, we assume the effect to be symmetric.} and calculated as $\Delta [X/H]_{avg} = (\Delta [X/H]_{upper} + \Delta [X/H]_{lower})/2$. The median \(\Delta \log(g)\) for main sequence stars is 0.12 dex, while for RGB stars the median gravity delta increases to 0.42 dex.

To derive the final uncertainty on the measurement, we combine the uncertainty from stellar parameter derivation as described above in quadrature with the uncertainties in the continuum fit and measurement of the line, both of which are provided by PySME when computing the [Mg/H] and [Si/H] measurements. The PySME errors provide an essential dependence on the S/N of the spectra, but are known to be overestimated \citep{wehrhahn_pysme_2023}. To resolve this we have rescaled our combined uncertainties for both [Mg/H] and [Si/H]. We divided the uncertainties by a factor defined as the ratio of the mean uncertainties for our stars with S/N between 100 and 200, and the mean uncertainty for APOGEE stars in the same S/N range with $\apomh < -2.3$. Adding an extra 0.05 dex error representing systematic uncertainties to these rescaled errors in quadrature gives a mean error of $\approx0.1$ for the high-S/N stars, while preserving the S/N dependence and the star-to-star variance. Finally, we add the factor $\rm |[Si/Mg]|/3$ to account for the uncertainty in the assumptions on [Mg/M] and [Si/M] described in Sect. \ref{sec:Mchoices}.

\subsection{Choice of the stellar parameters}
For the results presented in this work, we have adopted the stellar parameters (e.g. TEFF and LOGG) for the program stars from the APOGEE database, despite the lack of calibrated metallicities. To evaluate the robustness of the use of these parameters, we have also tried using SAGA database stellar parameters where available. Deriving metallicities with our pipeline yielded a mean offset of $\Delta$[M/H] = -0.02 dex with a standard deviation of 0.08 dex, with most stars being within one standard deviation of each other. These modest differences further support the use of the APOGEE parameters.

\section{Results}\label{sec:results}
\subsection{Extending APOGEE metallicities to the extremely metal-poor regime}

Out of the initial sample of 406 stars, our method converged to metallicities based on both [Mg/H] and [Si/H] or based on either of the two elements for 327 stars. Of these, 304 stars have an uncertainty on the final weighted metallicity below 0.3 dex. All results have been collected in Table \ref{tab:results}, with an explanation of each column in Table \ref{table:column_explainer}. We note that for stars with estimated metallicities below [M/H] $= -3.1$ dex, the lines would become too shallow and the errors increase rapidly, leading us to conclude that the metallicities below this value are better treated as upper limits.

\begin{figure}[]
    \centering
    \begin{subfigure}{0.47\textwidth}
        \centering
        \includegraphics[width=\linewidth]{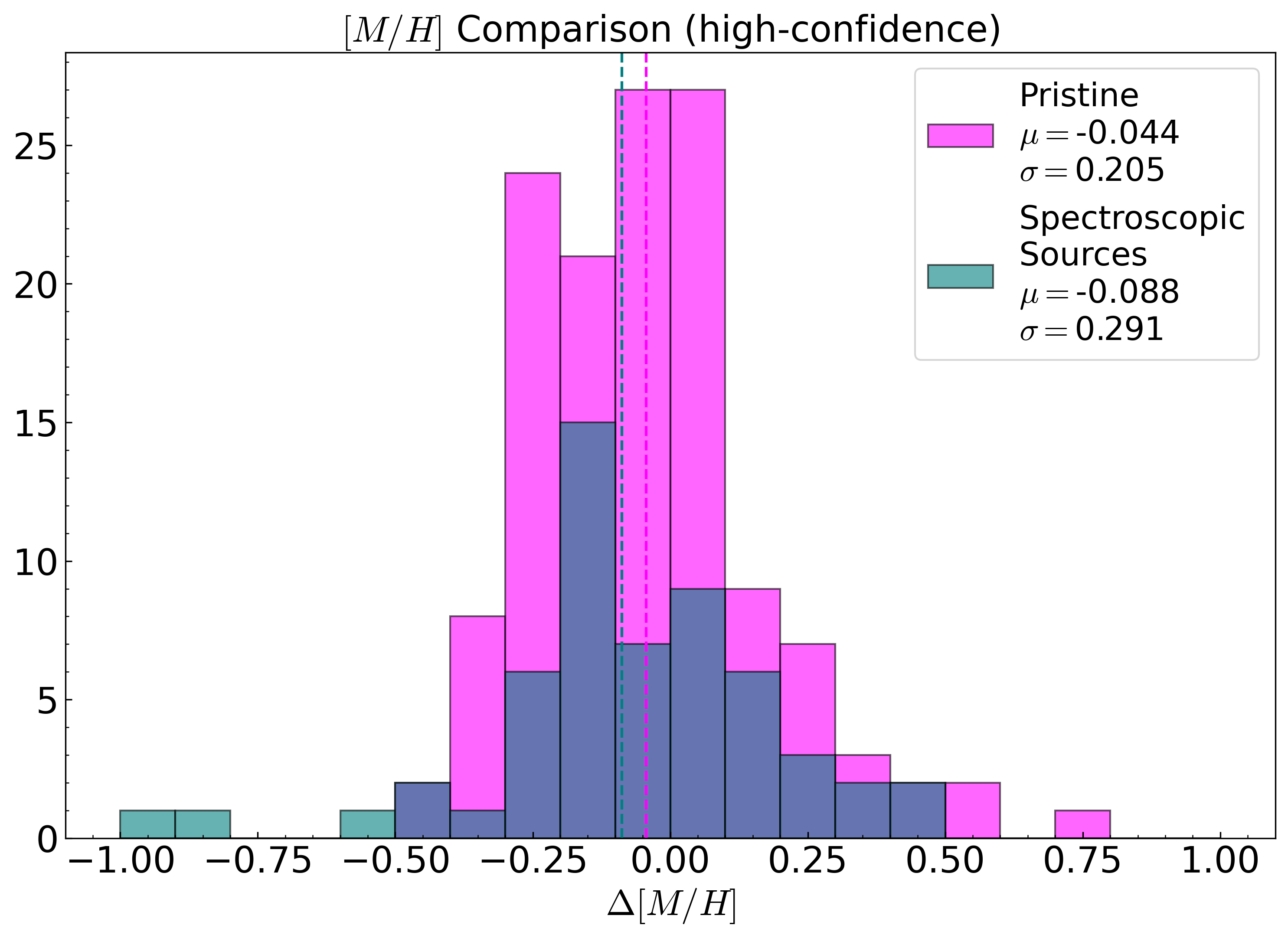}
    \end{subfigure}
    \caption{Residual plot for estimated metallicities against photometric and spectroscopic metallicities, shown as a histogram. The means are shown as dashed lines, and given along with the standard deviations. }
    \label{fig:comparisons}
\end{figure}

We validate our method making use of three groups of spectroscopic metallicities. Firstly, 29 stars in our sample for which high-resolution analysis exists in the SAGA database including at least temperature, gravity, and metallicity measurements. Secondly, we use the LAMOST DR10 reanalysis mentioned above, with 71 stars overlapping with our high-confidence sample. Finally we use the Ca II triplet based metallicities from \citealp[]{Viswanathan2024, Matsuno_Starkenburg_Balbinot_Helmi_2024} for 17 stars.
Additionally, 134 stars have photometric metallicity estimates from the Pristine Gaia-synthetic catalogue \citep[][in our selection we follow the quality cuts suggested in this work and we note that the E(B-V) values for these stars do place them within the trusted photometric metallicity regime - the median E(B-V) value of this sample is 0.04 with only four outliers with 0.3 $<$ E(B-V) $<$ 0.43]{martin_pristine_2023}. Figure \ref{fig:comparisons} shows histograms of the metallicity differences between our results and those of these external surveys, divided into spectroscopic values (SAGA, LAMOST, and data from \citealp[]{Viswanathan2024, Matsuno_Starkenburg_Balbinot_Helmi_2024}), and photometric metallicities from Pristine. 
No significant bias is observed with metallicity and the standard deviations are in range with expectations. We note that we did additionally check the correlation of the difference in metallicity as function of metallicity, temperature, and surface gravity for SAGA overlapping stars (16 stars) using the Pearson correlation coefficient, and while there is weak trend with metallicity (p-value: -0.69), no systematic trends were found with temperature or surface gravity. We have also done direct comparisons of our [Mg/H] and [Si/H] measurements to the ones in SAGA, shown in Fig \ref{fig:hi_res_MgSi_SAGA_comp} in the Appendix.
For the lower-confidence sample, only ten stars overlap with these reference catalogues. The agreement seems slightly worse, as perhaps expected, but also without any significant offsets (this is illustrated in Fig \ref{fig:comparisons_low} in the Appendix).

\begin{figure*}[]
  \centering
  \noindent 
\hspace*{-1.8cm}
\begin{subfigure}[t]{0.5\linewidth}
\vspace{12pt}
  \includegraphics[width=0.95\linewidth]{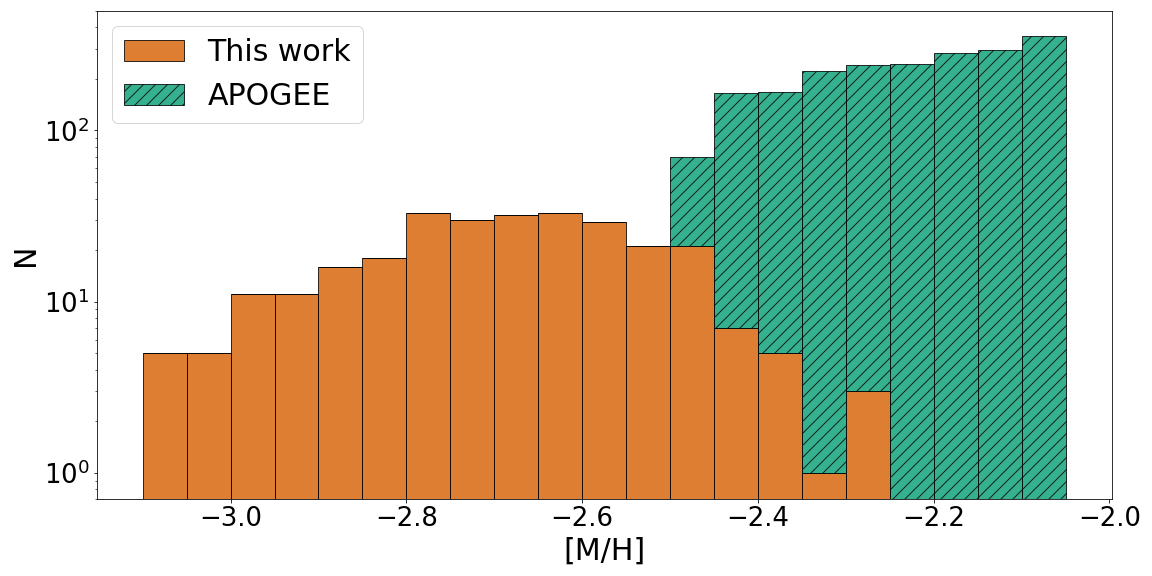}
\end{subfigure}
\hspace{-0.45cm} 
\begin{subfigure}[t]{0.5\linewidth}
\vspace{0pt}
  \includegraphics[width=1.2\linewidth]{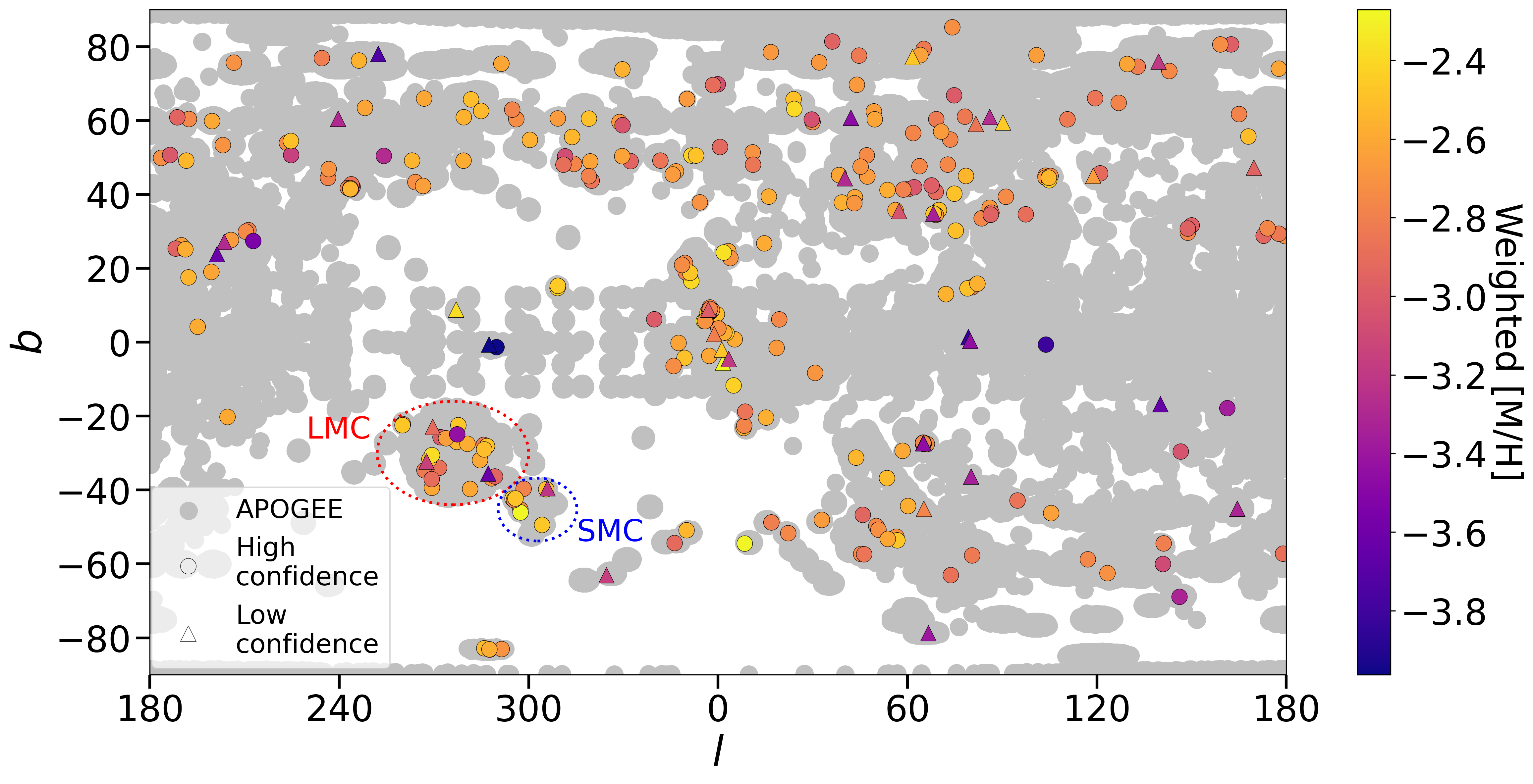}
\end{subfigure}

  \caption{The left panel is a histogram of the number of stars per bin (on a log scale) with stacked bars of the original truncated metal-poor tail of APOGEE with weighted metallicities from the high-confidence sample. The reported metallicities already include the quality cuts. On the right, the same clean sample with weighted metallicities is mapped onto the celestial sphere and superimposed on the APOGEE footprint. The location of the LMC and SMC, where there is a concentration of stars, has been marked.}
  \label{fig:mwMDF}
\end{figure*}

With the high-confidence sample, we extend the lowest end of APOGEE's metallicities to [M/H] = -3.1 dex. The resulting metallicity distribution function (MDF) is illustrated in the left panel of Fig. \ref{fig:mwMDF} as the number of stars in bins of 0.05 dex. Our sample has an average metallicity of [M/H] $= -2.7$ dex and nicely complements the distribution at the tail of APOGEE's original MDF. We also note that there are no catastrophic outliers identified in our analysis, all stars in the high-confidence sample have [M/H] $<  -2.25$ in our final spectroscopic analysis, and the stars in the lower-confidence sample all have [M/H] $< -1.9$.

\subsection{Special samples}

As already highlighted in Paper I, our sample contains some subsamples that are of special interest to the astronomical community. Due to its near-infrared wavelength range, APOGEE has a unique capability to look through dusty regions in the disc and bulge of the Milky Way. Due to the APOGEE targeting strategy, our sample of stars is bright, with mean G-magnitude of 14.4 mag, making them perfect targets for higher-resolution follow-up with relatively low exposure time. Additionally, in its targeting, it has deliberately included many known dwarf galaxies in its footprint, among which the Large and Small Magellanic Clouds (LMC and SMC, respectively). Paper I has already highlighted the stars that according to their positions and orbits belong to these structures. In this work, we validate their very metal-poor nature. Whenever a star is a member of any of these spatial or dynamically identified subsamples, this is highlighted in column Region in Table \ref{tab:results}.

An additional special subsample, highlighted in column EMP of Table \ref{tab:results} consists of stars with metallicities from this work lower than [M/H] $= -3$, also called extremely metal-poor stars. As noted above, some of these have extremely high uncertainties. We visually inspected these spectra in detail, and found that most had relatively high S/N and did not present any anomalies that would lead us to rule them out as unfit spectra. We therefore believe that the inability of our pipeline to measure more precise abundances is likely due to their true extremely metal-poor nature and that these stars are very interesting targets for follow-up studies.

\subsection{The ratio of Mg and Si}
While we assume that [Mg/M] = [Si/M] in our assessment of overall metallicities, this does not necessarily hold for all stars, especially very metal-poor stars can be born in very different environments \citep[e.g.][]{kobayashi_origin_2020}. As we measure both elements separately, we are in an excellent position to assess their abundance ratio in more detail. 
Because our [M/H] measurements are dependent on the assumed Mg and Si abundances, we will look at the evolution of [Si/Mg] as a function of [Mg/H]. 

For comparing [Si/Mg] ratios we have carefully selected stars from APOGEE DR17 and GALAH DR4 \citep{buder_galah_2024} with high S/N\footnote{For APOGEE we require S/N > 150. For GALAH we require S/N > 40 on all four CCDs, no flags on Mg and Si, and none of the 1, 4, 8, 9, 10 or 12 flags in sp\_flag.}, and stars from the SAGA database, excluding results from medium resolution (R $\leq 20000$) spectra. Due to anomalous [Si/Mg] ratios in globular clusters (GCs), we have used the membership probabilities from \cite{GCid2021MNRAS.505.5978V} to isolate GC stars. All stars with a probability $>0.9$ to belong to a globular cluster are removed from the following analysis and are further discussed in sections \ref{sec:SiMgCauses} and \ref{sec:GlobularClusters}. Dwarf galaxy stars have similarly been removed with membership probabilities from \cite{Battaglia2022DG}, requiring their (recommended) probability $>0.2$.

The main difference between these comparison samples is which wavelength region is observed. While our work, and APOGEE, uses near-infrared spectra, GALAH and SAGA are both based on optical spectra. There are several factors that can cause offsets between results from the optical and the near-infrared. The lines used for the analysis will necessarily be different, with different excitation energies and different NLTE effects. As we use the same spectra as APOGEE with the same atomic data for both Mg and Si lines this should not cause any offsets, but could be a contributing factor to offsets from the optically derived abundances. The APOGEE abundances have a zeropoint offset applied to align the abundances of solar neighbourhood stars with solar values. The zeropoint offset for [Si/Mg] is $-0.029$ dex \citep{APOGEEDR16ASPCAP_2020AJ....160..120J}\footnote{We cite the DR16 values as the offsets are not available for DR17.}.

\begin{figure}[h]
    \centering
    \begin{subfigure}[b]{  
    \textwidth}
        \includegraphics[width=0.47\textwidth]{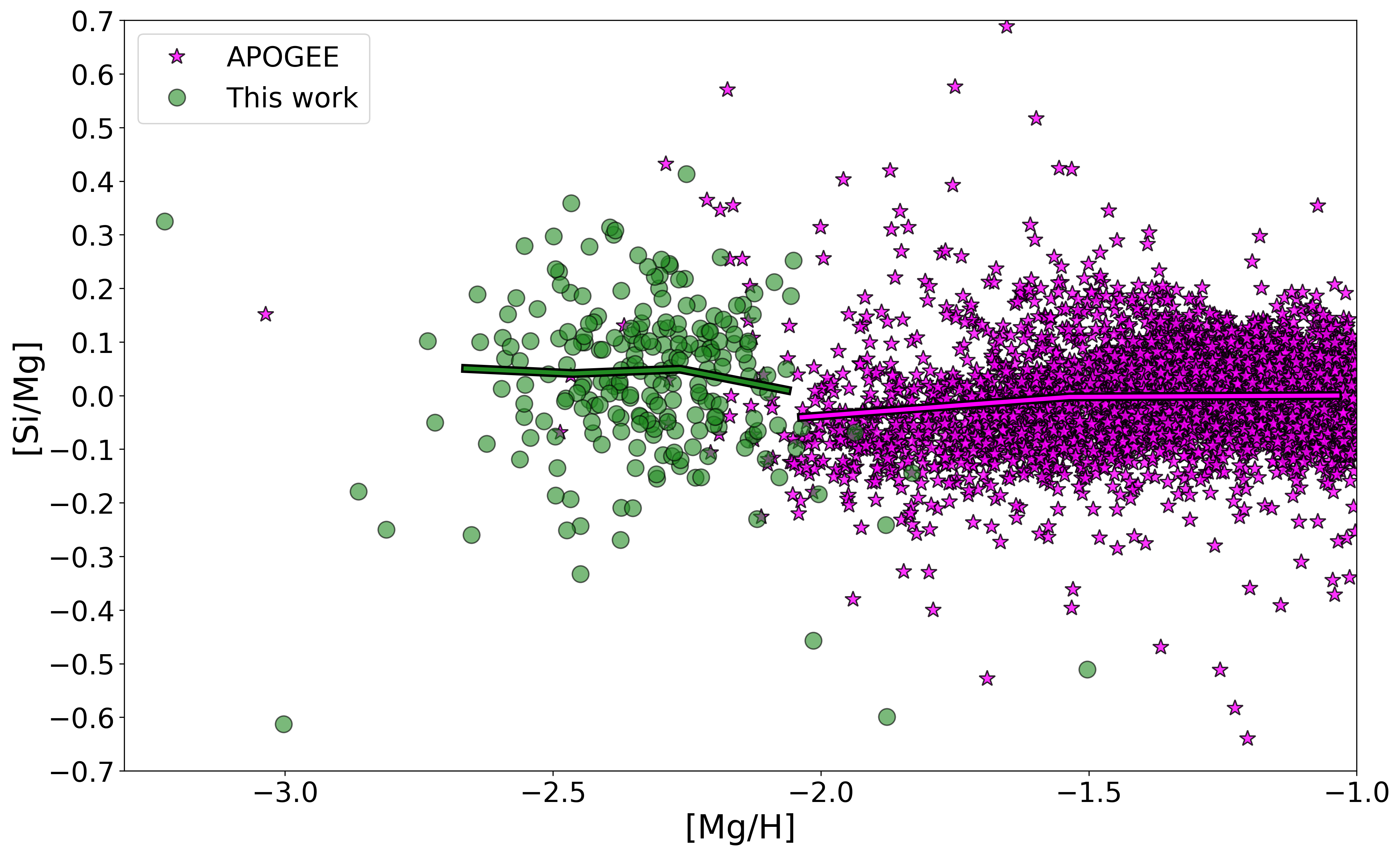}
    \end{subfigure}
    \begin{subfigure}[b]{
    \textwidth}
        \includegraphics[width=0.47\textwidth]{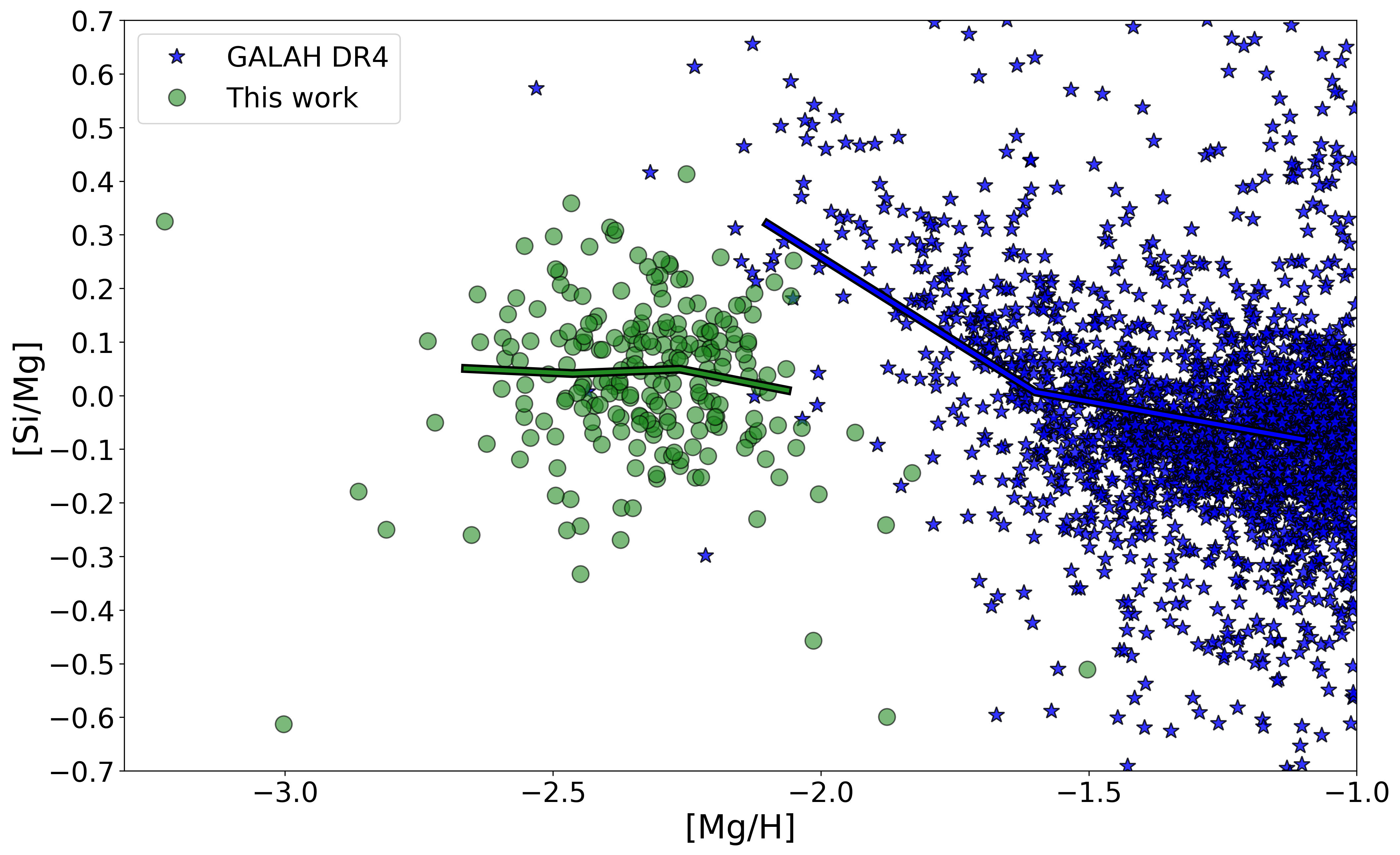}
    \end{subfigure}
    \begin{subfigure}[b]{
    \textwidth}
        \includegraphics[width=0.47\textwidth]{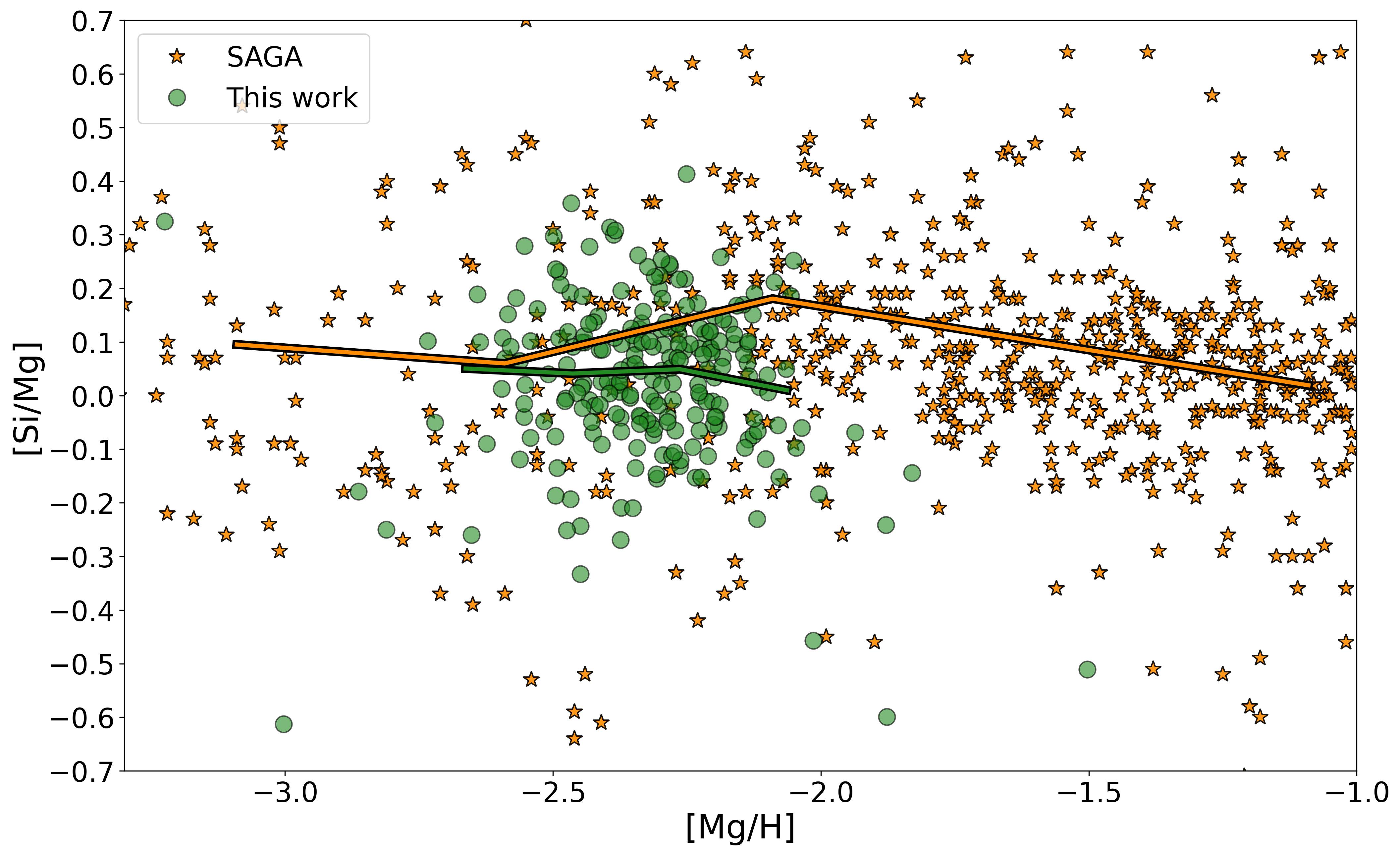}
    \end{subfigure}
    \caption{The three plots include the clean high-confidence sample, and all APOGEE, GALAH, and SAGA stars with [Mg/H] lower than -1. The APOGEE dataset has been filtered to exclude all stars from dwarf galaxies and GCs. The running mean shown for each dataset goes from the minimum to the maximum [Mg/H] with bins every 0.5 dex for the three surveys, and 0.2 dex for the high-confidence stars due to smaller sample size and spanning a smaller metallicity range.}
    \label{fig:MgSiSpace}
\end{figure}

The comparisons are shown in Fig. \ref{fig:MgSiSpace}. We note that our sample is located in a fairly unexplored region of this graph, with little overlap in [Mg/H] for APOGEE and GALAH, and significantly fewer stars in SAGA than at higher metallicities. Our sample thus provides new insights into this regime. In comparison to APOGEE, the mean [Si/Mg] of our sample is elevated by 0.033 dex. 
Correcting for the zeropoint offset in APOGEE would effectively eliminate the offset to our work, suggesting that our analysis does not introduce a significant bias compared to ASPCAP.
In contrast to this, our results are slightly enhanced compared to the GALAH abundances, with a mean offset of 0.073 dex. Similar to APOGEE, GALAH applies a zeropoint offset to their abundances. For [Si/Mg] this offset is $0.061$ dex \citep{buder_galah_2024}, which could account for the offset we observe. While appearing significantly more scattered, the mean from SAGA is offset in the other direction by $-0.060$ dex.
In light of these comparisons, it appears that our [Si/Mg] measurements are not significantly offset compared to optical values in this regime. The difference in offset between GALAH and SAGA could reflect the different methodologies of the samples, or the different selection functions.
For direct comparison of our [Mg/H] and [Si/H] abundances with values from SAGA, see Appendix \ref{app:saga_comp}.

\subsubsection{Globular clusters}
\label{sec:GlobularClusters}
\begin{figure}[h]
    \centering
    \includegraphics[width=1\linewidth]{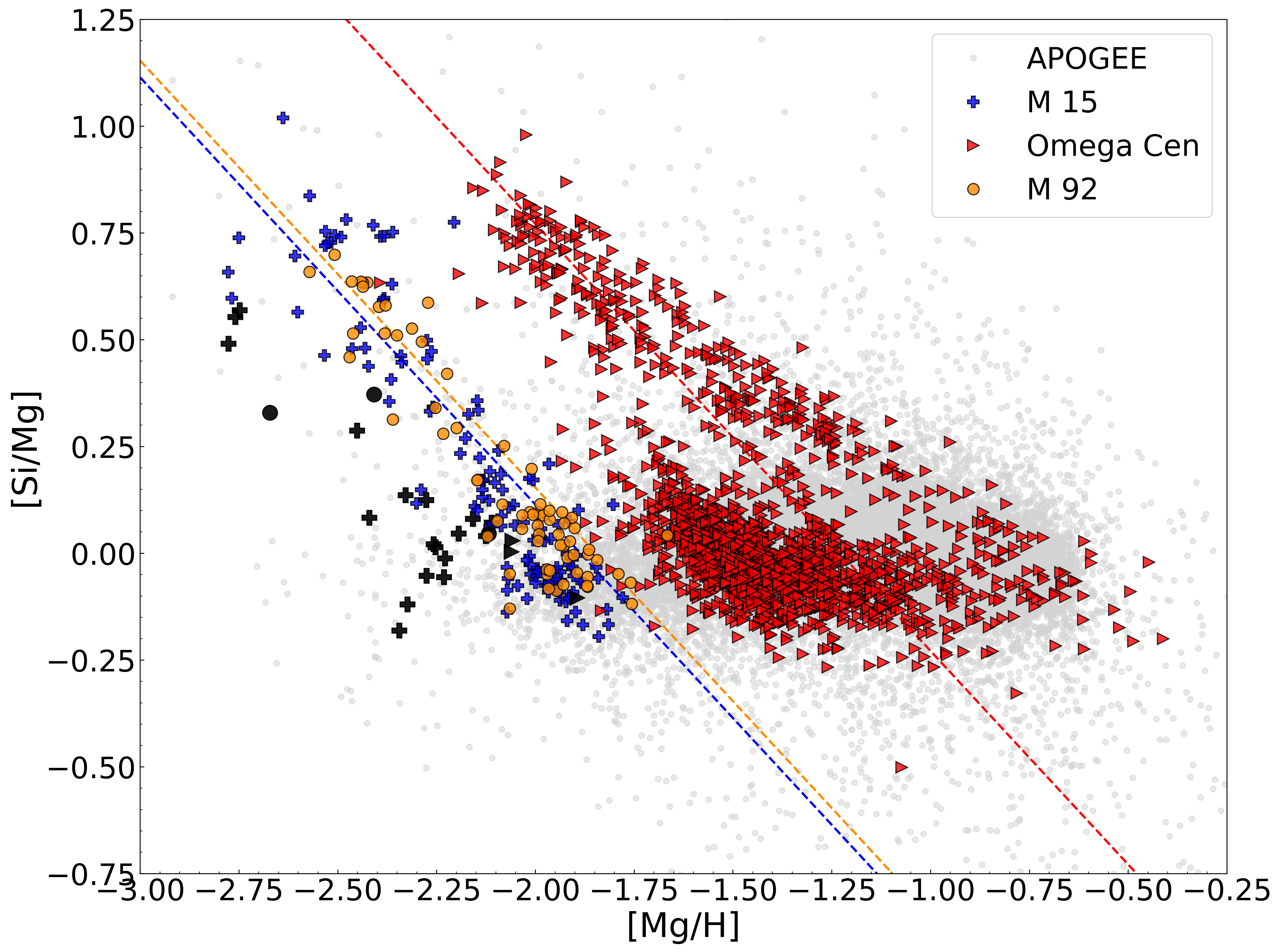}
    \caption{[Si/Mg] vs. [Mg/H] for the GC stars in our sample (shown in black), along with APOGEE abundances for the same GCs (shown with the same symbols as our measurements). The dotted lines indicate y=x lines for each cluster that match the cluster stars at high [Si/Mg] values.}
    \label{fig:GlobCluster}
\end{figure}

There are stars from three GCs present in our high-confidence sample: these are M 15, $\omega$ Cen and M 92.
Figure \ref{fig:GlobCluster} shows the distribution in [Si/Mg] versus [Mg/H]. We include in this figure stars that have a membership probability of at least 0.9 for any of these three clusters according to \citet{GCid2021MNRAS.505.5978V} from both our sample (black symbols) and the APOGEE catalogue (coloured symbols). We refer the reader to Fig \ref{fig:GlobClusterApp} in the Appendix for a compilation of other globular clusters (that have no stars from our dataset) in this abundance space in the APOGEE data.

Figure \ref{fig:GlobCluster} clearly shows that [Si/Mg] is not a constant for stars within a globular cluster system and that this ratio might be enhanced up to $\approx 1$ dex. Indeed, variations in the abundances of light elements for so called second population (2P) GC stars have been seen for several decades \citep[see e.g.][]{GC1994PASP..106..553K}. While the precise mechanism affecting the abundances of these stars is not clear, anti-correlations of C-N, Na-O and Mg-Al are commonly observed for 2P stars \citep[see][and references within]{GCreview2022Univ....8..359M}. A Mg-Si anti-correlation is rarer, but has been observed for some massive GCs. Even if Si is not enhanced, the depletion of Mg by the Mg-Al anti-correlation can make [Si/Mg] appear enhanced. We have therefore added diagonal lines to Fig. \ref{fig:GlobCluster}, any trend within a given globular cluster that is steeper than the line would thus indicate Si enhancement in addition to Mg depletion. 

Of the three GCs we observe, $\omega$ Cen displays the largest amount of chemical inhomogeneity, as it is well known for \citep[see e.g.][]{OmegaCen1975ApJ...201L..71F, OmegaCen1996ApJ...462..241N}. The split between populations is clear, with a largely flat group around [Si/Mg] $= 0$ and a diagonal trend with decreasing [Si/Mg] as [Mg/H] increases (consistent with what we would expect for 2P). For the other GCs, only the diagonal trend is distinct, with a minor density increase around [Si/Mg] $= 0$. While the trend for $\omega$ Cen is notably shallower than the one-to-one line, the lines for both M 15 and M 92 seem steeper than a purely diagonal line, hinting that in these clusters there is Si enhancement on top of Mg depletion. 

We subsequently focus our specific attention on the stars added to the APOGEE sample in this work. These stars are represented as black symbols in Fig. \ref{fig:GlobCluster} (their symbol shape reflects the cluster with which they are associated) and we quantify their mean [Si/Mg] and its dispersion for stars per globular clusters in Table \ref{tab:Mg_Si}. For $\omega$ Cen all three stars in our sample belong to the first group of stars with a flat distribution, and show a mean close to zero with a small spread. For M 15 and M 92 we have stars spread over the whole range of the distribution, matching well the slope of the stars already analysed within APOGEE. 

For more traditional methods of metallicity estimation, these GC stars could serve as a good validation sample. In our method there are two main issues making this difficult. The first one being Mg depletion in the 2P, which will make these stars appear to be more metal-poor. If Si enhancement is present it will offset this to some degree, but Si enhancement is less common. The second major effect is our selection function. Both M 15 and M 92 have metallicities just above the threshold of -2.5 that our method aims to select. This means that most stars in these GCs should be analysable by APOGEE. Any star that ends up in our selection must be on the lower end of the metallicity distribution, or appear to be so by defects in the spectra reducing the strength of the lines. The remaining GC, $\omega$ Cen, is known for its large metallicity distribution and is therefore a poor choice for calibration no matter the method.

\subsubsection{Dwarf galaxies}
\label{sec:DG_SiMg}

\begin{figure}
    \centering
    \includegraphics[width=1\linewidth]{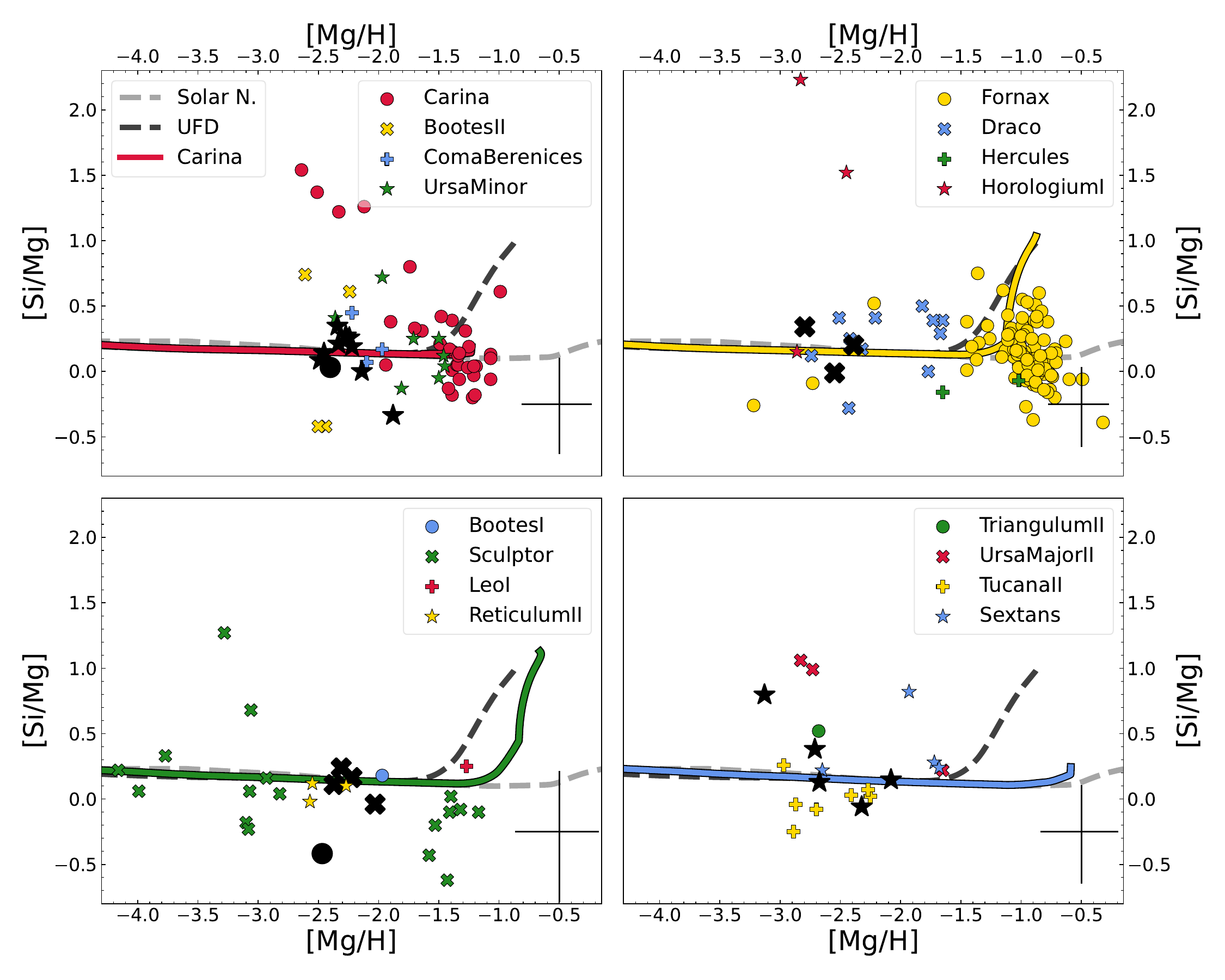}
    \caption{[Si/Mg] for stars from our sample in various dwarf galaxies (as identified by the criteria of \cite{Battaglia2022DG}; black symbols) supplemented with SAGA dwarf galaxy members according to the same criteria. Only measurements from high-resolution spectra are shown. Chemical evolution models for the Solar neighbourhood and a UFD are shown in all panels, with dedicated models for Carina, Fornax, Sculptor and Sextans shown in their respective panels.}
    \label{fig:SAGA_Si_Mg}
\end{figure}

To put the results for the dwarf galaxies into context, we will more closely examine the individual galaxies and compare them to literature values for other dwarf galaxies and to chemical evolution models for the Solar neighbourhood, Ultra Faint Dwarf (UFD) galaxies, and specialised models for Carina, Fornax, Sculptor and Sextans. The chemical evolution models we will use are the same as in \cite{2020ApJ...895..138K}, but with a metal-dependent delay-time distribution function of sub-Ch mass SNe Ia. For the solar neighbourhood, the evolution of $\alpha$ elements against Fe is very similar to the model in \cite{2020ApJ...895..138K}. As demonstrated in their Fig 39, with any contribution from SNe Ia, the [Si/Mg] ratio becomes higher. Although they do not affect [Si/Mg], Wolf-Rayet stars are also included as in \cite{2024ApJ...962L...6K}. The star formation histories are determined from independent observations, such as the observed metallicity distribution functions of these dwarf spheroidal galaxies. The UFD model is constructed by a compilation of stars in various UFDs, and slower star formation results in the rise of [Si/Mg] at a lower metallicity. Standard, Kroupa IMF is adopted in all models.

we show the same [Si/Mg] versus [Mg/H] abundance space in Fig. \ref{fig:SAGA_Si_Mg} for stars belonging to several dwarf galaxies (by crossmatching with \cite{Battaglia2022DG}, adopting the aforementioned probability cut of 0.2). We include stars from our work (black symbols) as well as dwarf galaxy stars from the SAGA database\footnote{Using data from \cite{SAGAsi2001ApJ...548..592S, SAGA2003AJ....125..684S, SAGAsi2004ApJ...612..447F, SAGA2004PASJ...56.1041S, SAGA2005AJ....129.1428G, SAGA2008ApJ...688L..13K, SAGAsi2009ApJ...701.1053C, SAGA2010A&A...523A..17L, SAGA2010ApJ...708..560F, SAGA2010A&A...524A..58T, SAGA2010ApJ...722L.104N, SAGAsi2012A&A...538A.100L, SAGAsi2012ApJ...751..102V, SAGA2013ApJ...763...61G, SAGA2014A&A...572A..88L, SAGA2014ApJ...786...74F, SAGAsi2015A&A...580A..18F, SAGA2015A&A...583A..67J, SAGA2015ApJ...802...93S, SAGA2015MNRAS.449..761U, SAGA2015ApJ...807..171J, SAGA2016ApJ...830...93J, SAGA2016ApJ...817...41J, SAGA2017ApJS..230...28N, SAGA2017ApJ...838...83K, SAGA2018ApJ...852...99N, SAGA2018ApJ...856..138J, SAGA2018ApJ...857...74C}. We use the same cuts to exclude studies based on medium resolution spectra that were used for Fig. \ref{fig:MgSiSpace}.}. The galaxies are distributed amongst the panels for legibility. 
We furthermore show chemical evolution models for the Solar neighbourhood and a typical Ultra Faint Dwarf (UFD) galaxy in each panel, along with a specialised models for Carina, Fornax, Sculptor and Sextans. The median uncertainty for the stars in each panel is also shown. 

The majority of the stars in Fig. \ref{fig:SAGA_Si_Mg} have [Si/Mg] ratios in the range -0.1 to 0.5, with our measurements aligning well with the SAGA data and generally following the behaviour expected by the models. A number of significant outliers with high [Si/Mg] can be seen for Carina, Horologium I and Sculptor, with [Si/Mg] > 1. Our highest measurement of [Si/Mg] $\sim$ 0.8 for a star in Sextans, is significantly lower than these outliers, and is in line with the abundances of a Sextans star from the literature. Our lowest outlier with [Si/Mg] $\sim$ -0.4 for a star in Bootes I, is amongst the lowest at this [Mg/H], and is significantly lower than the other star from Bootes I. A similar pattern can be seen for Bootes II, with large differences in [Si/Mg] for stars at similar [Mg/H].

Inclusion in Fig. \ref{fig:SAGA_Si_Mg} requires a measurement of both [Mg/H] and [Si/H]. As some of our stars only have one of the elements measured, they do not appear in the figure. One of the reasons for why an abundance could not be measured, is that the lines of that element were too weak to be detected above the noise level, possibly indicating a very low abundance. For this reason, it is possible that some of the most interesting outliers in this abundance space are not shown. For the dwarf galaxies, three stars are missing Si measurements, one each in Bootes I, Ursa Minor and Sextans. Six stars lack Mg measurements, one in Carina and five in Sextans. While follow-up observations would have to be made to confirm if they truly are outliers, the missing Si measurements from Bootes I and Ursa Minor align well with low [Si/Mg] stars found there, and likewise missing Mg measurements from Carina and Sextans align well with outliers found in SAGA and our measurements for these galaxies respectively. 

The dwarf galaxies are also included in Table \ref{tab:Mg_Si}. Their mean [Si/Mg] and associated dispersions show differences, but quantitative comparisons are made difficult by low sample sizes and differences in overall metallicity. Most notably, both Ursa Minor and Sextans show significantly higher mean values.

\begin{table}
\caption{Means and standard deviations of [Si/Mg] as measured in this paper for the kinematically selected groups from Paper I, and likely dwarf galaxy and globular cluster stars. }    
\label{tab:Mg_Si}    
\centering         
\begin{tabular}{c c c c}     
\hline\hline 
Region & $\mu$ & $\sigma$ & N\\
\hline
Halo & 0.02 & 0.15 & 159\\ \hline
Inner & 0.01 & 0.13 & 13\\ \hline
Outer & 0.09 & 0.11 & 11\\ \hline
Confined & 0.04 & 0.15 & 21\\ \hline
LMC & 0.04 & 0.15 & 9\\ \hline
SMC & 0.19 & 0.11 & 8\\ \hline
BootesI & -0.42 &  & 1\\ \hline
Carina & 0.03 &  & 1\\ \hline
Draco & 0.18 & 0.18 & 3\\ \hline
Sculptor & -0.09 & 0.23 & 3\\ \hline
Sextans & 0.23 & 0.32 & 6\\ \hline
UrsaMinor & 0.21 & 0.08 & 9\\ \hline
M 15 & 0.15 & 0.21 & 16\\ \hline
Omega Cen & -0.02 & 0.07 & 3\\ \hline
M 92 & 0.21 & 0.23 & 2\\ \hline
\end{tabular}
\end{table}

\subsubsection{Galactic environments}

\begin{figure}
    \centering
    \includegraphics[width=1\linewidth]{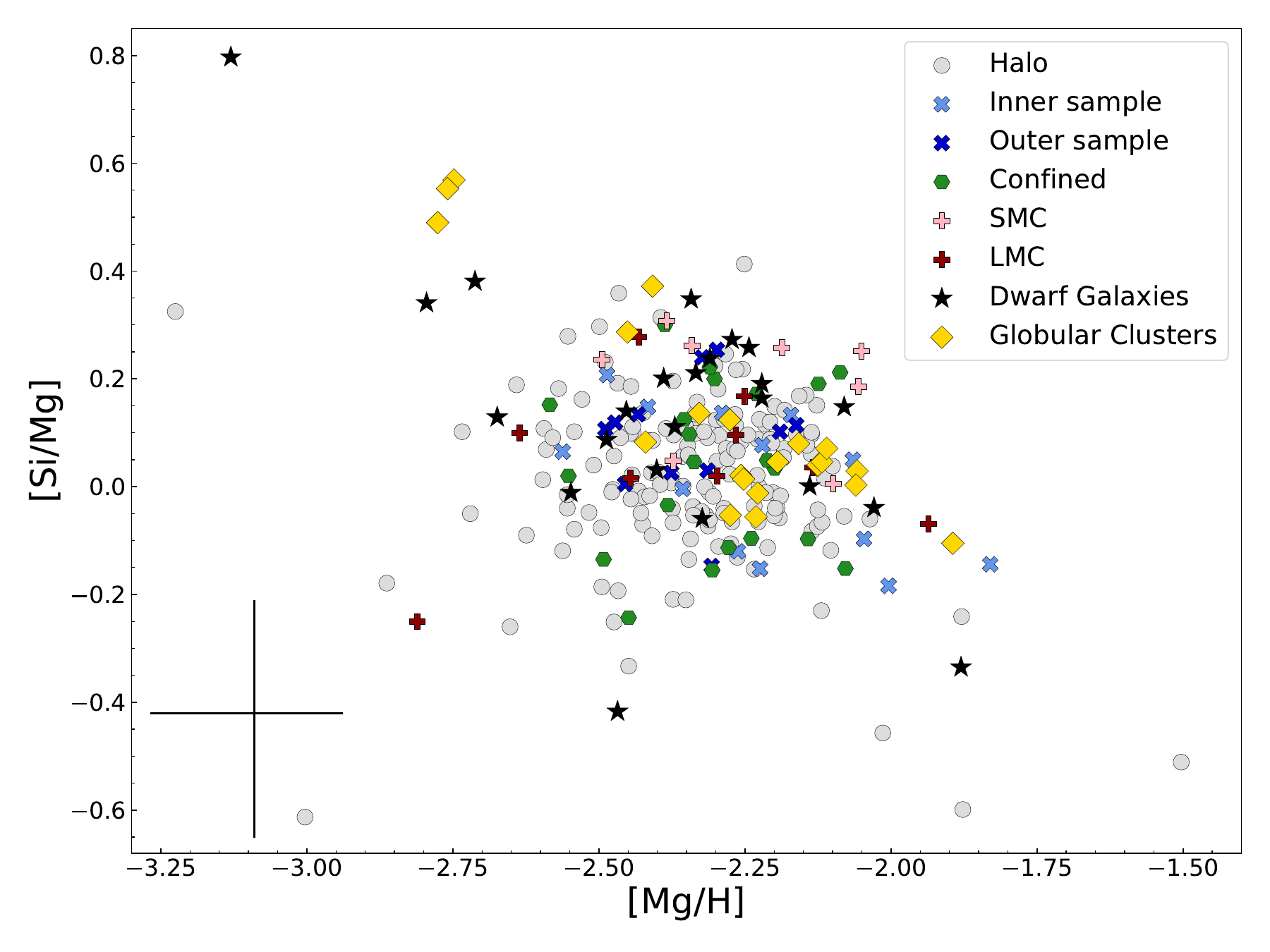}
    \caption{Abundances for the high-confidence sample, plotted as [Mg/H] against [Si/Mg]. The sample is split according to the kinematic selection made in Paper I, and also into different dwarf galaxies according to \cite{Battaglia2022DG} and for the Magellanic Clouds, the selection made in Paper I. The median errors for the abundances are shown in the bottom left corner.}
    \label{fig:Si_Mg_regions}
\end{figure}

To illustrate how the [Si/Mg] ratio changes for different regions of the Galaxy we show [Si/Mg] plotted against [Mg/H] in Fig. \ref{fig:Si_Mg_regions}. The figure shows only stars from the high-confidence sample of RGB stars. The stars are split up according to the orbital categories introduced by Paper I. These are: halo (Z$_{\rm max}$ > 3.5 kpc, Apo$_{\rm mean}$ > 3.5 kpc),  inner sample ( $R_{\rm GC}$ < 3.5 kpc, Apo$_{\rm mean}$ < 5 kpc), outer sample ( $R_{\rm GC}$ < 3.5 kpc, 5 < Apo$_{\rm mean}$ < 15 kpc), and confined stars (Z$_{\rm max}$ < 3.5 kpc, Apo$_{\rm mean}$ > 3.5). Additionally we show different symbols for stars likely belonging to (the outskirts of) the LMC and SMC (see for membership criteria Paper I) and the stars identified as belonging to dwarf galaxies by \cite{Battaglia2022DG} or globular clusters by \cite{GCid2021MNRAS.505.5978V}. 

The mean and standard deviation of the [Si/Mg] abundances and the number of stars associated with each of these kinematic selections are given in Table \ref{tab:Mg_Si}. Most notably, the mean of the SMC is offset one sigma above the general halo. The LMC is more similar to the halo, with a slightly larger dispersion. While the Inner sample is very similar to the Halo, the mean of the Outer sample is slightly higher but still within one sigma of the halo distribution. The confined sample is more similar to the inner sample, with no significant outliers. 

Generally speaking, for the kinematically defined inner, outer and confined samples, the sample sizes are too small and the uncertainties are too large to provide very clear statistics. However, we note that the slightly elevated [Si/Mg] in the outer sample could be consistent with a higher degree of accreted stars from substructures where we see enhancement (GCs and dwarf galaxies, as shown in the previous subsections).

\section{Discussion: Scenarios explaining the variation in [Si/Mg]}
\label{sec:SiMgCauses}
There are several mechanisms that can alter [Si/Mg]. In this section we will go over the most significant mechanisms and discuss where they could be signifiant for our measurements.

\subsection{Globular cluster origin}
For the stars in our sample that have been identified as members of GCs, the 2P light element anti-correlations offer a clear explanation to why some of them are enhanced in [Si/Mg]. For the stars with enhanced [Si/Mg] in the halo, or in dwarf galaxies, this opens up an intriguing scenario that some of them were born in a GC that has been stripped and phase-mixed. Such a hypothesis to explain these low-metallicity, high [Si/Mg] stars, is especially relevant in the present-day framework where more and more stellar streams of very low-metallicity clusters are discovered \citep[e.g.][]{WanPhoenix20, MartinC1922}, indicating that (lower mass) cluster formation was more efficient in the early Universe than previously assumed. We have investigated whether we could confirm such an enrichment hypothesis for some of our halo stars by crossmatching our results with \textsc{STREAMFINDER} \citep{streamfinder2024ApJ...967...89I}, a catalogue of stellar stream candidates based on Gaia data. While two of our stars overlap, both match with the Orphan stream, that is not associated with a GC progenitor. Additionally, both these stars are not enhanced in [Si/Mg].

As mentioned above, it is possible that stars associated with dwarf galaxies are also enhanced by the GC 2P anti-correlation. As the [Si/Mg] enhancement does not require Si enhancement, merely Mg depletion, it is not limited to the rarer GCs with the Si-Mg anti-correlation. Additional exploration of other elements often found in anti-correlations in globular clusters (C, N, Na, O, and Al) would be helpful to further investigate this possibility by constraining Mg–Al, Si–Al, Ca–Al, and O–Mg correlations. If a sufficient number of enriched stars could be associated within a dwarf galaxy, the slope of their enhancement in the [Si/Mg] versus [Mg/H] space could potentially give information on the nature of that origin.

\subsection{Supernovae Ia}
As chemical enrichment proceeds more slowly in dwarf galaxies compared to the Milky Way, the knee in the [$\alpha$/Fe] vs [Fe/H] distribution, indicating the onset of SNe Ia, will occur at lower metallicities \citep{deBoer2014, Lemasle2014A&A...572A..88L}. The majority of our stars have [Mg/H] lower than the threshold for SNe Ia contribution for their galaxies. As such, we do not expect enhancement in our measurements to be caused by SNe Ia. The SAGA stars shown in Fig. \ref{fig:SAGA_Si_Mg} cover a wider range of [Mg/H], and cross the knee for several of the dwarf galaxies, most significantly for Fornax (at [Mg/H] $\sim-1.5$ \citep{hendricks_metal-poor_2014}) and  Carina (at [Mg/H] $\sim-2.1$ \citep{CarinaKnee2014A&A...572A..10D}). As Si is produced significantly more in SNe Ia compared to Mg, an upward trend at higher [Mg/H] is expected \citep{KobayashiIa2020ApJ...895..138K}. This could explain the dispersion in [Si/Mg] for Fornax at the highest metallicities. 

\subsection{Supernovae II}
While SNe II are believed to be the source of most of the Mg and Si in the stars in our sample, there are differences in how much of each element is produced depending on the mass and explosion energies of the SNe. More massive stars are predicted to produce different fractions of Mg and Si compared to lower-mass stars, but the models can vary significantly in their quantitative predictions \citep{kobayashi_galactic_2006, Heger2010}. Additionally, not all of it is deposited in the ISM.  
Hypernovae, with a high explosion energy, release more of the Si compared to faint or fall-back supernovae \citep[e.g.][]{Heger2010}, where the heavier Si is retained in the SNe remnant to a higher degree. The shape of the top of the IMF will therefore be important for the [Si/Mg] ratio. For smaller systems where only a few massive SNe are likely to have contributed to the earliest star formation, the sampling of the IMF could introduce anomalous abundances. Our outlier at low [Si/Mg] in Bootes I, would be consistent with the low hypernovae contribution deduced for the galaxy from the low [Ba/Fe] ratio. Low [Si/Mg] ratios may be caused by aspherical (jet-like) explosions \citep{nomoto_chemical_2008}. Similar to fallback SNe this mechanism can explain very low [Si/Mg] ratios.
Conversely, the high [Si/Mg] outlier in Sextans aligns well with the enhanced [Sr/Ba] found in Sextans stars \citep{MashonkinaSextans2022}, which could be connected to hypernovae \citep{IzutaniHypernova2009, Mashonkina2017A&A...608A..89M}. Another signature of hypernovae enrichment would be enhancement in Zn. In the SAGA database, five of the dwarf galaxy stars in our sample have Zn measurements\footnote{Using data from \cite{SAGAsi2012ApJ...751..102V, SAGA2013ApJ...763...61G, SAGA2004PASJ...56.1041S, SAGAzn2010ApJ...719..931C, SAGAzn2017A&A...606A..71S}.}, but neither our Bootes I nor the Sextans outliers have it available. One of the stars do appear enhanced in [Zn/Fe], our second star in Bootes I ($\rm [Zn/Fe] = 0.4$). We do not have a [Si/H] measurement for this star due to artifacts in the spectra close to the Si lines, but Si does not appear to be enhanced. 

As the number of significant outliers in the dwarf galaxies is small, another plausible interpretation to the outliers is inhomogeneous mixing of SNe ejecta in the ISM. As SNe with different progenitor masses will produce different ratios of elements, and in a dwarf galaxy with a more limited number of progenitors, this could be a source of outliers.
This has been suggested previously for [Ca/Mg] variations in Carina by \cite{SAGAsi2012ApJ...751..102V} based on simulations by \cite{Revaz2012A&A...538A..82R}. 

For the Magellanic Clouds, the evolution of $\alpha$ elements have been the topic of some attention in recent years \citep{alpha_MC2012ApJ...761..180B, alpha_MC2020ApJ...895...88N}, but the specific ratio of [Si/Mg] is not considered as frequently. The analysis of APOGEE stars by \cite{alpha_APOGEE_abunds2021ApJ...923..172H} hints at a higher [Si/Fe] ratio for the SMC compared to the LMC at higher metallicities, but the results for the SMC also appear more scattered. \cite{alpha_APOGEE_model2024ApJ...974..227H} study the [Si/Mg] versus [Mg/H] space for the Magellanic Clouds with APOGEE data, but are constrained to the more metal-rich end ([Mg/H] > -2). Their APOGEE selection indicates a slightly elevated [Si/Mg] for dwarf galaxies as a group at [Mg/H] < -1, but the values for the individual dwarf galaxies are not shown. They also demonstrate with chemical evolution models how a top-light IMF without the most massive stars could increase the [Si/Mg] ratio \citep[see Fig. 13][]{alpha_APOGEE_model2024ApJ...974..227H}. 
Studies independent of APOGEE also find some level of [Si/Mg] enhancement in the Clouds, using young stars \citep{alpha_MC_Astar1999ApJ...518..405V}, the interstellar medium \citep{alpha_MC_ISM2024A&A...683A.216D} and optical spectroscopy \citep{LMC_abunds2013A&A...560A..44V}.

\subsection{Pair-instability supernova}

High [Si/Mg] ratios ([Si/Mg] > 0.5) have also been identified as a possible Pair-instability supernova (PISN) signature \citep{PISNe2018ApJ...857..111T}. However, additional abundances are required to confirm whether PISN as a likely progenitor. One of the elements that could be important for this is aluminium, which has strong lines in the H band and thus provides an alternative to sodium in the optical wavelength range (in the same vein, calcium can be used in the optical to trace silicon). The [Al/Mg] ratio expected for PISN descendants is very low ([Al/Mg]<-1), which makes it very unlikely to be measured from APOGEE spectra. We indeed verified that no Al line is detected for any of these spectra, and no meaningful upper limits could be derived either. If the stars observed with elevated [Si/Mg] would have a strong PISN contribution to their abundance pattern, we would furthermore expect: solar [O/Mg], extremely low nitrogen levels, abundance ratios of [C/Mg], [Na/Mg], and [Al/Mg] all approximately –1. Iron abundances can further constrain the progenitor mass and rotating PISN are predicted to differ from their non-rotating counterparts in their nitrogen yields: N levels become comparable to those of C and O. A very characteristic element for PISN is Zn, with extremely low yields from low-mass models and low [Zn/Fe] or [Zn/Mg] ratios ranging from –1 to significantly lower values. 
Of the five stars with Zn measurements in SAGA mentioned above, all have $\rm [Si/Mg]< 0.2$ from our measurements and none are depleted in Zn, with [Zn/Fe] = 0.10 being the lowest value. We note that \citet{sestito_extended_2023} carefully examined the possible contribution of PISN in Ursa Minor, one of the dwarf galaxies that shows a general higher [Si/Mg] trend, but concluded it was ruled out. The outlier star in Sextans does meet the [Si/Mg] > 0.5 criteria, but has no recorded measurements of other elements to verify a PISN signature. 

\section{Conclusions and future outlook}
We have reanalysed a sample of 327 very metal-poor stars from the APOGEE spectrum database that had no calibrated metallicity estimates from the general pipeline. We demonstrated that it is possible to measure elemental abundances in the infrared for stars with a metallicity lower than -2.5 in the H band. While in the first paper of this series, Paper I, a method was designed to create a reliable sample of very metal-poor stars from APOGEE and investigate their orbits, in this work, we have designed a specialised pipeline for analysing the remaining available Mg and Si lines at these low abundances. We used these lines to measure Mg and Si abundances as well as a general metallicity for our sample, and we confirm their very metal-poor nature. 

The [Si/Mg] ratios we derived highlight some intriguing differences between stars of different populations. The SMC stars have significantly higher [Si/Mg] than the halo, as is also the case for stars in Ursa Minor. Other interesting stars are highlighted among the highly likely members of various dwarf galaxies. As seen in the GCs, light element anti-correlations can be a method of achieving high [Si/Mg] enhancement, even with minimum Si enhancement. Their occurrence might also be related to the ratio of hypernovae (or even pair-instability supernovae) products that enriched these galaxies at an early stage. Our most interesting stars for testing these theories are found in Sextans (high [Si/Mg]) and Bootes I (low [Si/Mg]).

The ratio of [Si/Mg] alone is not sufficient to precisely constrain the origin of a star, and several scenarios have been explored for many cases in this work. Especially in stars exhibiting enhanced [Si/Mg], follow-up observations and a determination of more abundance ratios in the same stars are needed to correctly determine progenitor scenarios.

The analysis we presented might be very useful for future H-band spectroscopic surveys, such as the MOONS instrument that is soon to be deployed at the VLT \citep{cirasuolo_moons_2020}. In addition, our analysis provides precise abundances for Si at metallicities lower than can be measured by most current optical surveys. This adds another important $\alpha$-element to the abundance ratio analysis.

\section*{Data availability}
Table \ref{tab:results} is available in electronic form at the CDS via anonymous ftp to cdsarc.u-strasbg.fr (130.79.128.5) or via http://cdsweb.u-strasbg.fr/cgi-bin/qcat?J/A+A/.

\begin{acknowledgements}

M.M. and E.S. acknowledge funding through VIDI grant ``Pushing Galactic Archaeology to its limits'' VI.Vidi.193.093,  which is funded by the Dutch Research Council (NWO). This research has been partially funded from a Spinoza award by NWO (SPI 78-411). This research was supported by the International Space Science Institute (ISSI) in Bern, through ISSI International Team project 540 (The Early Milky Way). This work has received funding from the European Research Council (ERC) under the Horizon Europe research and innovation programme (Acronym: EARLYMW, Grant number: 101170507). Funding for the Sloan Digital Sky Survey IV has been provided by the Alfred P. Sloan Foundation, the U.S. Department of Energy Office of Science, and the Participating Institutions. SDSS acknowledges support and resources from the Center for High-Performance Computing at the University of Utah. The SDSS web site is www.sdss4.org.
SDSS is managed by the Astrophysical Research Consortium for the Participating Institutions of the SDSS Collaboration including the Brazilian Participation Group, the Carnegie Institution for Science, Carnegie Mellon University, Center for Astrophysics | Harvard \& Smithsonian (CfA), the Chilean Participation Group, the French Participation Group, Instituto de Astrofísica de Canarias, The Johns Hopkins University, Kavli Institute for the Physics and Mathematics of the Universe (IPMU) / University of Tokyo, the Korean Participation Group, Lawrence Berkeley National Laboratory, Leibniz Institut für Astrophysik Potsdam (AIP), Max-Planck-Institut für Astronomie (MPIA Heidelberg), Max-Planck-Institut für Astrophysik (MPA Garching), Max-Planck-Institut für Extraterrestrische Physik (MPE), National Astronomical Observatories of China, New Mexico State University, New York University, University of Notre Dame, Observatório Nacional / MCTI, The Ohio State University, Pennsylvania State University, Shanghai Astronomical Observatory, United Kingdom Participation Group, Universidad Nacional Autónoma de México, University of Arizona, University of Colorado Boulder, University of Oxford, University of Portsmouth, University of Utah, University of Virginia, University of Washington, University of Wisconsin, Vanderbilt University, and Yale University.
Guoshoujing Telescope (the Large Sky Area Multi-Object Fiber
Spectroscopic Telescope LAMOST) is a National Major Scientific
Project built by the Chinese Academy of Sciences. Funding for the project has been provided by the National Development and Reform Commission. LAMOST is operated and managed by the National Astronomical Observatories, Chinese Academy of Sciences.
This research was supported by the International Space Science Institute (ISSI) in Bern, through ISSI International Team project 540 (The Early Milky Way).
This work has made use of the VALD database, operated at Uppsala University, the Institute of Astronomy RAS in Moscow, and the University of Vienna.
\end{acknowledgements}

\bibliographystyle{aa} 
\bibliography{MScPostThesis_250606} 
\newpage

\appendix

\section{Results table}
The results of our spectroscopic analysis are presented in Table \ref{tab:results}. Details on the columns are shown in Table \ref{table:column_explainer}. Measurements for both the high- and low-confidence samples are included, as well as the stars that have been removed by visual inspection.

\begin{sidewaystable*}
\caption{Extract of the full table of results to be made available electronically. Descriptions of the columns are presented in Table \ref{table:column_explainer}.}  
\label{tab:results}   
\centering                       
\begin{tabular}{c c c c c c c c c c c c c c c}  
\hline\hline
        source\_id & ra & dec & TEFF & LOGG & Flag & Sample & S/N & M\_H & M\_H err& Mg\_H & Si\_H & Lit. spec & Region & EMP \\ 
        \ & [deg] & [deg] & [K] & [dex] & \ & \ & \ & [dex] & [dex] & [dex] & [dex] & \ & \ & \ \\ \hline
1206440580982968704 & 242.2381 & 22.5140 & 4900$\pm$ 26 & 1.8$\pm$ 0.1 & 3  & 1 & 94 & -2.6 & 0.11 & -2.3$^{+0.10}_{-0.12}$ & -2.2$^{+0.05}_{-0.09}$ &   & H & \ \\ \hline
1095977904781400320 & 115.2676 & 67.1338 & 5320$\pm$ 39 & 2.9$\pm$ 0.1 & 3  & 1 & 94 & -2.8 & 0.12 & -2.3$^{+0.14}_{-0.16}$ & -2.4$^{+0.09}_{-0.10}$ & * & H & \ \\ \hline
2608992211966631168 & 338.2448 & -10.1329 & 5124$\pm$ 35 & 2.6$\pm$ 0.1 & 3  & 1 & 94 & -2.6 & 0.13 & -2.3$^{+0.10}_{-0.11}$ & -2.1$^{+0.12}_{-0.14}$ &   & H & \ \\ \hline
3650528932367561472 & 221.3746 & -1.2552 & 5087$\pm$ 33 & 2.5$\pm$ 0.1 & 3  & 1 & 93 & -2.4 & 0.16 & -2.2$^{+0.22}_{-0.26}$ & -2.0$^{+0.11}_{-0.13}$ &   & H & \ \\ \hline
1645415547490608384 & 228.3763 & 67.1102 & 4531$\pm$ 20 & 1.0$\pm$ 0.1 & 3  & 1 & 93 & -2.5 & 0.23 & -2.2$^{+0.24}_{-0.24}$ & -2.0$^{+0.25}_{-0.24}$ & * & UrsaMinor & \ \\ \hline
1770025090051944448 & 323.3399 & 12.7636 & 5113$\pm$ 38 & 2.5$\pm$ 0.1 & 3  & 1 & 93 & -2.7 & 0.14 & -2.2$^{+0.15}_{-0.18}$ & -2.4$^{+0.10}_{-0.12}$ &   & H & \ \\ \hline
1745949083942421504 & 322.4083 & 12.1995 & 4836$\pm$ 25 & 1.8$\pm$ 0.1 & 3  & 1 & 92 & -2.6 & 0.10 & -2.3$^{+0.19}_{-0.23}$ & -2.2$^{+0.09}_{-0.11}$ &   & M 15 & \ \\ \hline
3887764470323417600 & 153.1543 & 13.7081 & 4881$\pm$ 26 & 1.6$\pm$ 0.1 & 3  & 1 & 92 & -3.2 & 0.13 & -2.7$^{+0.19}_{-0.23}$ & -2.8$^{+0.13}_{-0.15}$ & * & H & \ \\ \hline
1398482862437904256 & 240.5563 & 46.0700 & 5182$\pm$ 38 & 2.7$\pm$ 0.1 & 3  & 1 & 91 & -2.8 & 0.12 & -2.3$^{+0.15}_{-0.17}$ & -2.4$^{+0.10}_{-0.11}$ &   & C & \ \\ \hline
4108901513082802304 & 257.0177 & -26.8994 & 4857$\pm$ 26 & 1.8$\pm$ 0.1 & 3  & 1 & 90 & -2.5 & 0.10 & -2.0$^{+0.11}_{-0.14}$ & -2.1$^{+0.07}_{-0.09}$ & * & I & \ \\ \hline
2613732481471351424 & 329.9184 & -11.2870 & 5094$\pm$ 37 & 2.5$\pm$ 0.1 & 3  & 1 & 90 & -2.9 & 0.16 & -2.5$^{+0.18}_{-0.22}$ & -2.6$^{+0.18}_{-0.20}$ &   & H & \ \\ \hline
3829054779943345536 & 153.4240 & -2.1901 & 4529$\pm$ 21 & 0.8$\pm$ 0.1 & 3  & 1 & 90 & -2.8 & 0.29 & -2.7$^{+0.30}_{-0.36}$ & -2.3$^{+0.18}_{-0.19}$ & * & Sextans & \ \\ \hline
4112211184855035392 & 256.0187 & -25.8176 & 4775$\pm$ 24 & 1.5$\pm$ 0.1 & 3  & 1 & 89 & -2.7 & 0.12 & -2.3$^{+0.09}_{-0.15}$ & -2.4$^{+0.09}_{-0.13}$ & * & I & \ \\ \hline
1738839195076542592 & 320.5891 & 6.3290 & 4903$\pm$ 30 & 1.8$\pm$ 0.1 & 3  & 1 & 89 & -2.6 & 0.10 & -2.2$^{+0.19}_{-0.23}$ & -2.2$^{+0.09}_{-0.12}$ &   & H & \ \\ \hline
1096118607909233408 & 118.1415 & 67.1204 & 5345$\pm$ 53 & 3.3$\pm$ 0.1 & 3  & 1 & 89 & -3.0 & 0.19 & -2.6$^{+0.23}_{-0.27}$ & -2.6$^{+0.25}_{-0.27}$ &   & H & \ \\ \hline
5178632322954178560 & 39.7226 & -6.5777 & 4811$\pm$ 25 & 1.5$\pm$ 0.1 & 3  & 1 & 88 & -2.9 & 0.16 & -2.5$^{+0.15}_{-0.26}$ & -2.4$^{+0.11}_{-0.21}$ &   & H & \ \\ \hline
1424320389257881856 & 244.0036 & 50.5439 & 4867$\pm$ 26 & 1.8$\pm$ 0.1 & 3  & 1 & 88 & -2.5 & 0.09 & -2.1$^{+0.08}_{-0.10}$ & -2.2$^{+0.08}_{-0.11}$ & * & H & \ \\ \hline
4111995921113976320 & 257.0354 & -25.7340 & 4922$\pm$ 28 & 2.0$\pm$ 0.1 & 3  & 1 & 88 & -2.7 & 0.17 & -2.4$^{+0.21}_{-0.25}$ & -2.2$^{+0.14}_{-0.16}$ & * & C & \ \\ \hline
3951071364849361408 & 180.7090 & 19.0618 & 4593$\pm$ 21 & 0.9$\pm$ 0.1 & 3  & 1 & 87 & -2.6 & 0.13 & -2.2$^{+0.15}_{-0.17}$ & -2.1$^{+0.17}_{-0.18}$ &   & H & \ \\ \hline
3628485957613518080 & 199.0081 & -6.8123 & 4772$\pm$ 25 & 1.5$\pm$ 0.1 & 3  & 1 & 86 & -2.5 & 0.13 & -2.1$^{+0.08}_{-0.13}$ & -2.2$^{+0.16}_{-0.18}$ &   & H & \ \\ \hline
1261551024344105600 & 225.0830 & 21.7677 & 5300$\pm$ 45 & 2.9$\pm$ 0.1 & 3  & 1 & 86 & -3.1 & 0.25 & -2.5$^{+0.29}_{-0.35}$ & -2.7$^{+0.19}_{-0.21}$ &   & H & \ \\ \hline
5972285340875232000 & 258.2171 & -39.4903 & 4921$\pm$ 28 & 1.9$\pm$ 0.1 & 3  & 1 & 86 & -2.6 & 0.09 & -2.2$^{+0.13}_{-0.15}$ & -2.2$^{+0.10}_{-0.12}$ &   & \ & \ \\ \hline
131045813048526976 & 36.3466 & 28.9178 & 4943$\pm$ 31 & 1.8$\pm$ 0.1 & 1  & 1 & 85 & -3.0 & 0.17 & -2.6$^{+0.11}_{-0.13}$ & nan & * & H & \ \\ \hline
4703711675634898816 & 10.5992 & -67.6494 & 4532$\pm$ 21 & 0.7$\pm$ 0.1 & 3  & 1 & 84 & -2.5 & 0.22 & -2.2$^{+0.17}_{-0.20}$ & -1.9$^{+0.19}_{-0.20}$ &   & SMC & \ \\ \hline
3608743053810741504 & 198.4250 & -14.4735 & 4924$\pm$ 29 & 1.7$\pm$ 0.1 & 3  & 1 & 84 & -2.9 & 0.10 & -2.5$^{+0.22}_{-0.27}$ & -2.5$^{+0.10}_{-0.12}$ & * & H & \ \\ \hline
6049700103459502464 & 243.7338 & -24.5012 & 4996$\pm$ 30 & 1.8$\pm$ 0.1 & 3  & 1 & 83 & -2.5 & 0.11 & -2.0$^{+0.13}_{-0.16}$ & -2.1$^{+0.10}_{-0.13}$ &   & H & \ \\ \hline
3686374351462686336 & 200.7340 & -1.3600 & 4530$\pm$ 21 & 0.8$\pm$ 0.1 & 3  & 1 & 83 & -2.5 & 0.14 & -2.1$^{+0.14}_{-0.17}$ & -2.1$^{+0.20}_{-0.21}$ & * & H & \ \\ \hline
1433128989225211392 & 259.4107 & 58.0775 & 4749$\pm$ 27 & 1.4$\pm$ 0.1 & 3  & 1 & 82 & -2.9 & 0.19 & -2.8$^{+0.51}_{-0.62}$ & -2.5$^{+0.07}_{-0.09}$ & * & Draco & \ \\ \hline
2478219120752687872 & 21.0915 & -7.5586 & 4987$\pm$ 30 & 1.9$\pm$ 0.1 & 3  & 1 & 81 & -3.3 & 0.40 & -3.2$^{+1.15}_{-1.38}$ & -2.9$^{+0.28}_{-0.31}$ &   & H & * \\ \hline
1301924850798604288 & 243.5225 & 23.5025 & 5616$\pm$ 50 & 2.4$\pm$ 0.1 & 3  & 2 & 80 & -2.7 & 0.23 & -2.5$^{+0.18}_{-0.22}$ & -2.2$^{+0.18}_{-0.23}$ &   & \ & \ \\ \hline
4061570977806163712 & 262.9129 & -26.8637 & 4579$\pm$ 22 & 1.0$\pm$ 0.1 & 3  & 1 & 80 & -2.7 & 0.18 & -2.4$^{+0.27}_{-0.30}$ & -2.3$^{+0.16}_{-0.15}$ &   & O & \ \\ \hline
1433158676039064192 & 260.4925 & 57.9345 & 4530$\pm$ 21 & 1.0$\pm$ 0.1 & 3  & 1 & 78 & -2.7 & 0.18 & -2.4$^{+0.18}_{-0.15}$ & -2.2$^{+0.17}_{-0.13}$ & * & Draco & \ \\ \hline
4107573234311253376 & 258.2238 & -29.1121 & 5023$\pm$ 32 & 2.0$\pm$ 0.1 & 3  & 1 & 78 & -2.9 & 0.20 & -2.6$^{+0.26}_{-0.25}$ & -2.4$^{+0.21}_{-0.16}$ & * & C & \ \\ \hline
6327510957665689728 & 221.5805 & -7.7394 & 4902$\pm$ 30 & 2.0$\pm$ 0.1 & 3  & 1 & 78 & -2.7 & 0.19 & -2.3$^{+0.11}_{-0.13}$ & -2.1$^{+0.21}_{-0.23}$ &   & O & \ \\ \hline
1645194511293583616 & 229.3008 & 66.7776 & 4545$\pm$ 23 & 0.9$\pm$ 0.1 & 3  & 1 & 78 & -2.7 & 0.19 & -2.3$^{+0.12}_{-0.15}$ & -2.1$^{+0.23}_{-0.25}$ &   & UrsaMinor & \ \\ \hline
3641322652789440256 & 217.2128 & -6.2453 & 5160$\pm$ 28 & 2.4$\pm$ 0.1 & 3  & 1 & 77 & -2.9 & 0.15 & -2.4$^{+0.11}_{-0.13}$ & -2.6$^{+0.11}_{-0.12}$ &   & H & \ \\ \hline
4651129593635024384 & 81.7586 & -73.3484 & 4433$\pm$ 20 & 0.7$\pm$ 0.1 & 3  & 1 & 77 & -2.6 & 0.24 & -2.3$^{+0.30}_{-0.36}$ & -2.2$^{+0.25}_{-0.28}$ &   & LMC & \ \\ \hline
4629265426946935040 & 61.0086 & -74.2750 & 4848$\pm$ 27 & 1.5$\pm$ 0.1 & 3  & 1 & 76 & -2.7 & 0.14 & -2.3$^{+0.12}_{-0.23}$ & -2.3$^{+0.17}_{-0.24}$ &   & LMC & \ \\ \hline
2772003641237241600 & 0.5039 & 14.9268 & 5158$\pm$ 37 & 2.1$\pm$ 0.1 & 3  & 1 & 76 & -2.6 & 0.09 & -2.3$^{+0.08}_{-0.09}$ & -2.2$^{+0.10}_{-0.12}$ &   & H & \ \\ \hline
4032024523351835904 & 176.8643 & 35.4369 & 5100$\pm$ 42 & 2.5$\pm$ 0.1 & 3  & 1 & 75 & -2.6 & 0.10 & -2.3$^{+0.13}_{-0.15}$ & -2.2$^{+0.08}_{-0.09}$ &   & H & \ \\ \hline
    \end{tabular}
\end{sidewaystable*}

\begin{table*}
\caption{Definitions for the columns in Table \ref{tab:results}.}             
\label{table:column_explainer}      
\centering          
\begin{tabular}{ c c }  
\hline\hline       
Column name & details \\ 
\hline                    
   source\_id & Gaia DR3 Source ID\\
   RA & Right accension in degrees, from APOGEE\\
   DEC & Declination in degrees, from APOGEE\\
   TEFF & Calibrated effective temperature, from APOGEE\\
   TEFF\_ERR & Uncertainty for the calibrated effective temperature, from APOGEE\\
   LOGG & Calibrated surface gravity, from APOGEE\\
   LOGG\_ERR & Uncertainty for the calibrated surface gravity, from APOGEE\\
   Flag & \makecell{Visual inspection flag: 0, unanalysable spectrum; 1, useable Mg; \\2, useable Si; 3, useable Mg and Si}\\
   Sample & Confidence sample flag: 0, unanalysable spectrum; 1, high-confidence sample; 2, low confidence sample\\
   S/N & Signal-to-noise ratio measured by APOGEE, per pixel\\
   M\_H & Weighted [M/H] as calculated in Sect. \ref{sec:flowchart}\\
   M\_H\_unc & Weighted uncertainty for the weighted [M/H]\\
   Mg\_H & [Mg/H] as calculated in Sect. \ref{sec:flowchart}\\
   Mg\_H\_lower & Lower uncertainty estimate for [Mg/H]\\
   Mg\_H\_upper & Upper uncertainty estimate for [Mg/H]\\
   Si\_H & [Si/H] as calculated in Sect. \ref{sec:flowchart}\\
   Si\_H\_lower & Lower uncertainty estimate for [Si/H]\\
   Si\_H\_upper & Upper uncertainty estimate for [Si/H]\\
   Lit. spec & \makecell{Stars marked * have a spectroscopic metallicity from the literature sources used in Paper I,\\ \citealp[]{Viswanathan2024, Matsuno_Starkenburg_Balbinot_Helmi_2024, buder_galah_2024, suda_stellar_2008}\\ \citealp[]{DESIDR1, PIGSII2020MNRAS.496.4964A, SEGUE_trashcan2016}\\ \citealp[]{Aguado2019MNRAS.490.2241A, LAMOST2012, simbad2000AJ....120.1351S, simbad2012PASP..124..519F, simbad2014ApJ...784..170X}\\ \citealp[]{simbad2014ApJ...794...60T, simbad2015MNRAS.448.2717W, simbad2018AJ....156..257S, simbad2020AA...641A.127R, simbad2020ApJ...894...34D}\\ \citealp[]{simbad2021ApJ...913...11L, simbad2022ApJ...926...26S, simbad2022MNRAS.513.3993O, simbad2023MNRAS.524..577O}}\\
   Region & \makecell{Region of the Local Group the star is assigned to in Paper I: H, halo; I, inner sample $R_{\rm GC}<3.5$ kpc Apo$_{\rm mean} < 5$ kpc;\\ O, outer sample $R_{\rm GC}<3.5$ kpc $5$ kpc < Apo$_{\rm mean} < 15$ kpc; C, confined sample $z_{\rm max}<3.5$ kpc Apo$_{\rm mean} > 3.5$ kpc;\\ LMC, Large Magellanic Cloud; SMC, Small Magellanic Cloud; other dwarf galaxies have their names written out}\\
   EMP & Extremely metal-poor star candidates \\
   
\hline                  
\end{tabular}
\end{table*}

\section{Additional comparison to literature abundances}
\label{app:saga_comp}
\begin{figure}
    \centering
    \begin{subfigure}{0.45\textwidth}
        \centering
        \includegraphics[width=\linewidth]{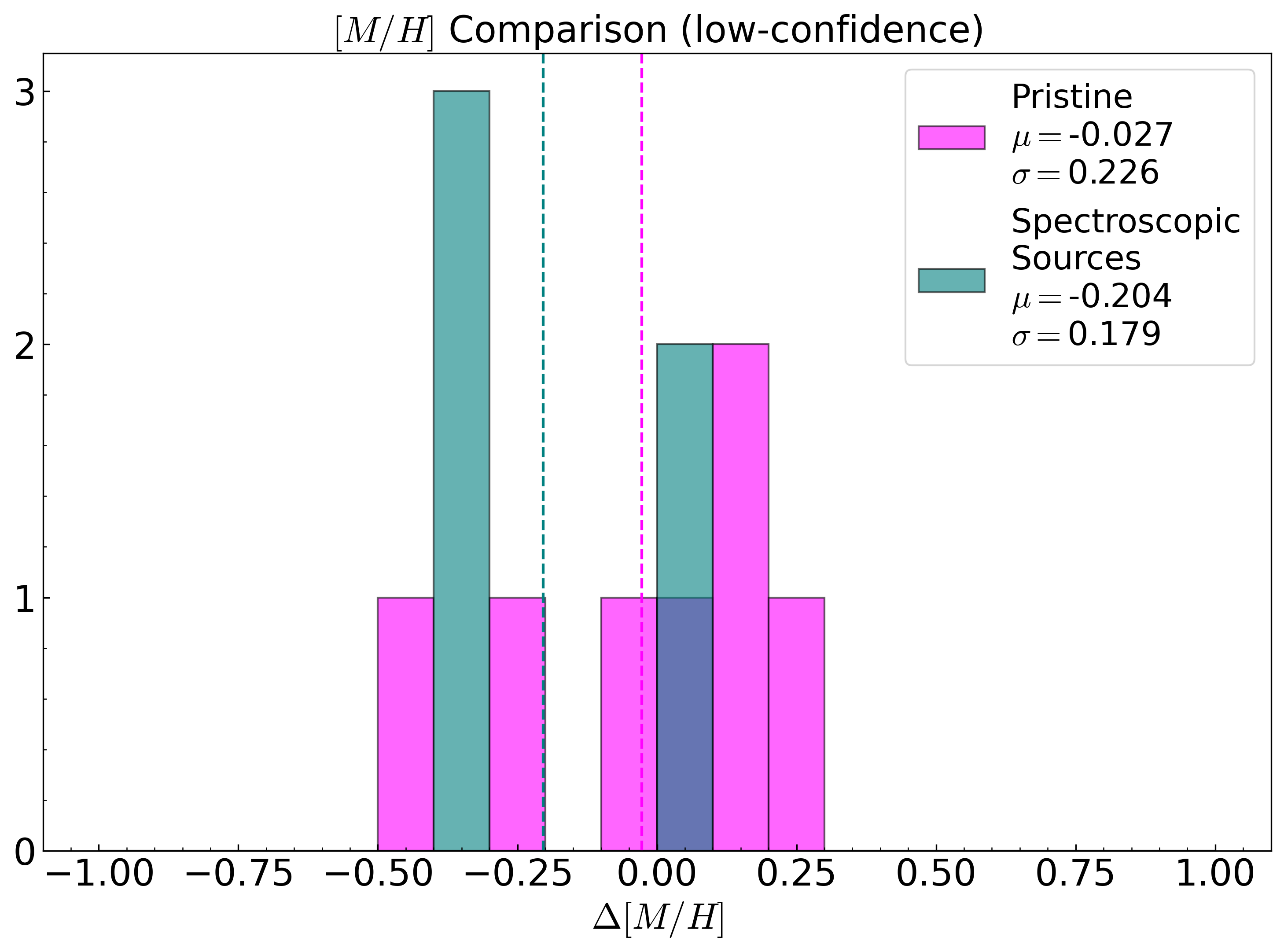}
    \end{subfigure}

    \caption{Analogous figure to Fig. \ref{fig:comparisons}, but for the lower-confidence sample.}
    \label{fig:comparisons_low}
\end{figure}

While the lower-confidence sample has significantly fewer overlaps with metallicity measurements from the literature, there are some that can be shown. In Fig. \ref{fig:comparisons_low} we show a histogram of the differences between our metallicities and the literature metallicity values from Pristine and spectroscopic sources for the lower-confidence sample, analogous to Fig. \ref{fig:comparisons}. Due to the very low number statistics, it is difficult to make direct comparisons between the samples, especially as the Kiel diagram is not evenly sampled for the few stars we do have literature values for in the lower-confidence sample. As there are no catastrophic outliers, we can conclude that the lower-confidence sample can be used to find interesting stars, but with higher uncertainty.

\begin{figure}
    \centering
    \includegraphics[width=1\linewidth]{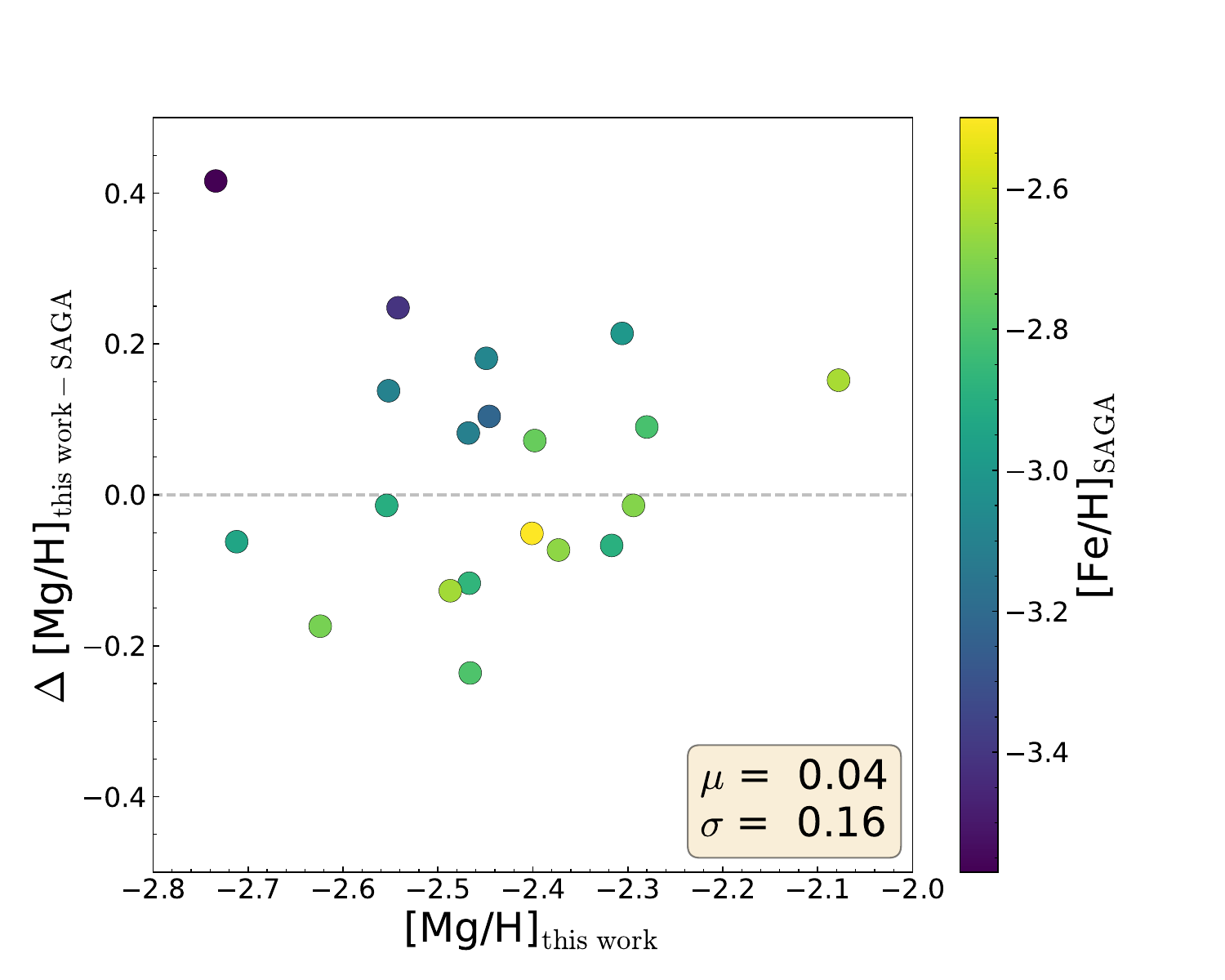}
    \includegraphics[width=1\linewidth]{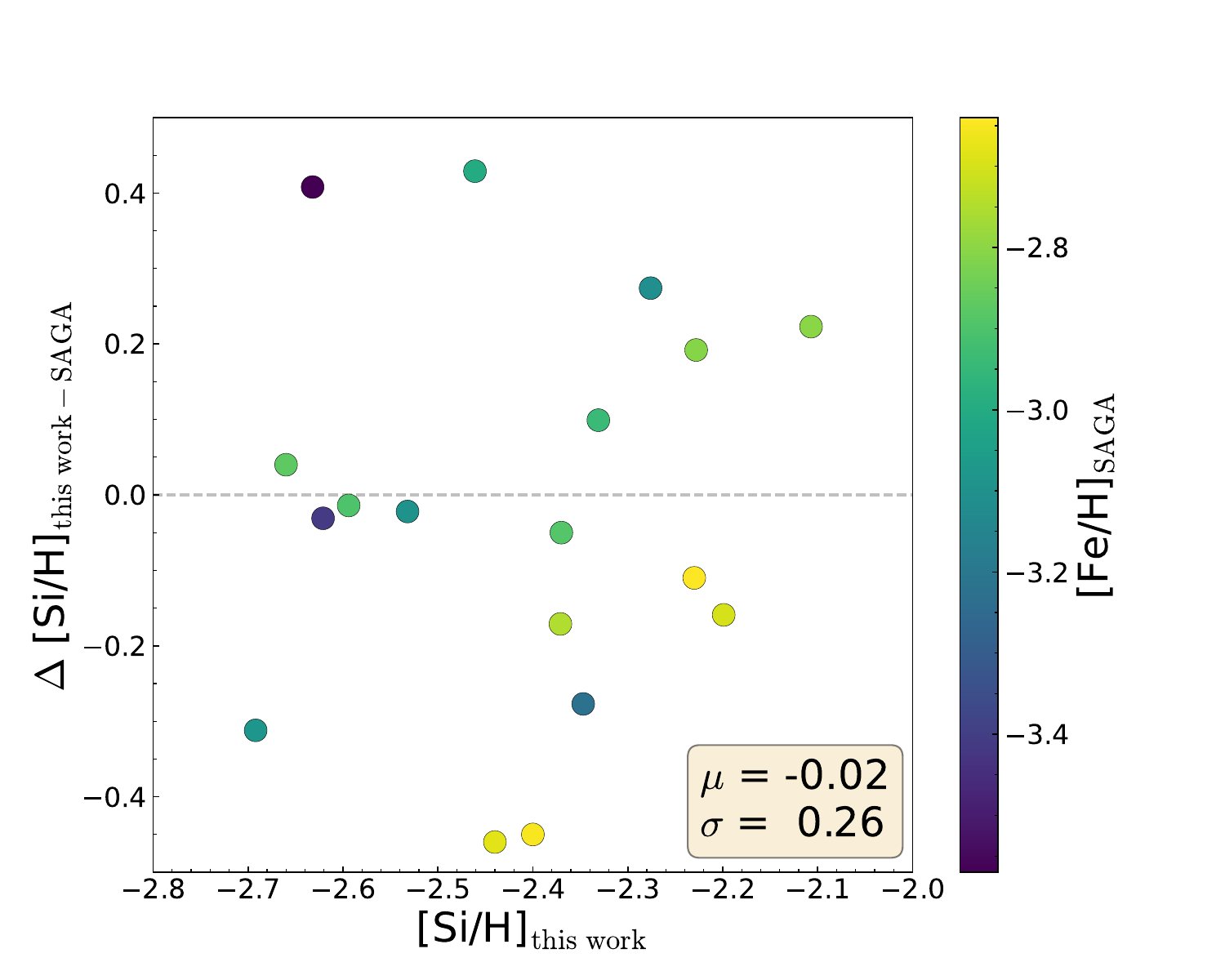}
    \caption{Direct comparisons of [Mg/H] (top panel) and [Si/H] (bottom panel) from this work and from the same SAGA sample used for Fig. \ref{fig:SAGA_Si_Mg}, with results derived from medium resolution spectra removed. The colourbar shows $\feh$ as compiled by SAGA.} 
    \label{fig:hi_res_MgSi_SAGA_comp}
\end{figure}

As we mainly focus on the [Si/Mg] ratio in Sect. \ref{sec:results}, we show a comparison between our [Mg/H] and [Si/H] measurements with those from SAGA in Fig. \ref{fig:hi_res_MgSi_SAGA_comp}, colour-coded by the $\feh$ from SAGA. For both elements the mean offset is small ($\mu_{Mg}=0.04$ and $\mu_{Si}=-0.02$), indicating that there are no significant systematic offsets between our near-infrared abundances and the optical ones from SAGA. The standard deviations are of the same order of magnitude as the standard deviation observed for the metallicity ($\sigma_{Mg}=0.16$ and $\sigma_{Si}=0.26$). The significantly lower value for $\sigma_{Mg}$ could be attributed to either a higher precision in our [Mg/H] values or to a lack of precision in SAGA [Si/H] values. As there are three Mg lines compared to the two Si lines, and the strongest Si line suffers from a blend with a hydrogen line, a higher precision for the [Mg/H] measurements would be expected. As for the SAGA [Si/H], Si is more difficult to measure from optical spectra, compared to Mg. However, as we exclude the results derived from medium resolution spectra, we judge it unlikely that the SAGA [Si/H] values are the main cause of the observed difference in $\sigma$.

\section{[Si/Mg] in globular clusters from APOGEE}
\begin{figure*}[h]
    \centering
    \includegraphics[width=0.75\linewidth]{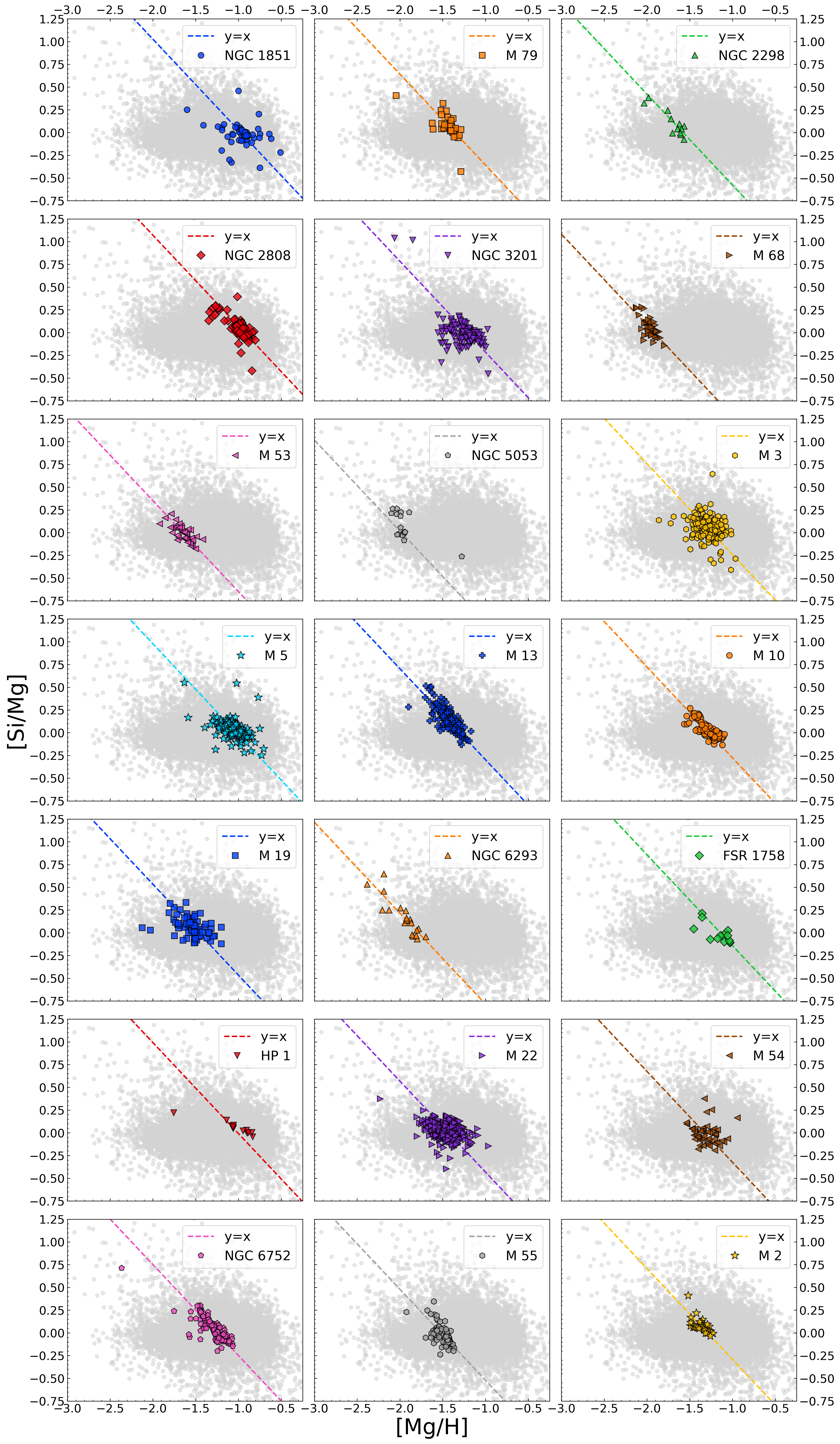}
    \caption{[Si/Mg] versus [Mg/H] for globular clusters identified by \cite{GCid2021MNRAS.505.5978V}, following Fig. \ref{fig:GlobCluster}. As we do not have stars from the high-confidence sample in any of these clusters, the abundances shown are from APOGEE, using the same sample shown in Fig. \ref{fig:MgSiSpace}.}
    \label{fig:GlobClusterApp}
\end{figure*}

In Fig. \ref{fig:GlobCluster} we show [Si/Mg] versus [Mg/H] for the three GCs with which we can associate some of our stars. For a wider view of how GCs behave in this chemical space we present the full crossmatch between APOGEE and the GC identifications from \cite{GCid2021MNRAS.505.5978V} in Fig. \ref{fig:GlobClusterApp}. Similar to the Figure above, we have added lines corresponding to y=x to assist with spotting Si enhancement. 

\end{document}